\documentclass{article}
\usepackage[utf8]{inputenc}
\usepackage{soulutf8,longtable,colortbl,setspace,ifthen,pdflscape}
\usepackage{adjustbox}
\usepackage{titlesec}
\usepackage{algorithm}
\usepackage{amsmath}
\usepackage{amsfonts}
\usepackage{amssymb}
\usepackage[english]{babel}
\usepackage{balance}
\usepackage{booktabs}
\usepackage{caption}
\usepackage{cellspace}
\usepackage{color}
\usepackage{comment}
\usepackage{adjustbox}
\usepackage{diagbox}
\usepackage{enumitem}
\usepackage{epsfig}
\usepackage{fancybox}
\usepackage{float}
\usepackage[T1]{fontenc}
\usepackage{footnote}
\usepackage{graphicx}
\usepackage{hyperref, cleveref}
\usepackage[utf8]{inputenc}
\usepackage{listings}
\usepackage{marvosym}
\usepackage{multirow}
\usepackage{pdflscape}
\usepackage{subcaption}
\usepackage{tablefootnote}
\usepackage[flushleft]{threeparttable}
\usepackage{url}
\usepackage{xr}
\usepackage{xspace}
\usepackage{pdfpages}
\usepackage{amsmath}
\newcommand{\al}{\textit{et al.}}
\newcommand{\ie}{{\textit{i.e., }}}
\newcommand{\eg}{{\textit{e.g., }}}
\graphicspath{{img/}}

\titleformat{\paragraph}
{\normalfont\normalsize\bfseries}{\theparagraph}{1em}{}
\titlespacing*{\paragraph}
{0pt}{3.25ex plus 1ex minus .2ex}{1.5ex plus .2ex}

\author{Amine Barrak}
\date{November 2020}

\title{RESEARCH PROPOSAL \\ 
\vspace*{30px}
\textbf{Toward a traceable, explainable, and fair JD/Resume recommendation system}\vspace*{40px}}

\author{\vspace*{20px}
	\textbf{Amine Barrak} \\ \\
	Supervised By: \\ \\
	\begin{tabular}{l}
        \large{Professor \textsc{Amal Zouaq} (Research Supervisor)}\\ 
        \large{Professor \textsc{Bram Adams} (Research Supervisor)}\\
    \end{tabular}
    \vspace*{80px}
	\\
	\\
	
	Department of Computer and Software Engineering \\
	Polytechnique Montréal, Québec, Canada
}
\date{March 2021}

\newcommand{\ThesisHyp}{(1) study the state-of-the-art in the JD/Resume matching ; (2) explore the performance of training an accurate JD/Resume models by combining a knowledge base with modern language models for recommendation purposes; 
(3) provide an explainable report to the stakeholders of the matching decisions recommendations; (4) adapt a traceable existing model to be able to track the different layers of the proposed matching and explainable architecture}
\newcommand{\RQOne}{What is the state-of-the-art in JD/Resume matching?}

\newcommand{\RQTwo}{Can knowledge base and modern language models improve JD/Resume matching?}

\newcommand{\RQFour}{How explain the decision of JD/Resume matching to concerned stakeholders?}
\newcommand{\RQFive}{Can traceable models be integrated into a JD/Resume matching process with low impact on the system complexity?}
\begin{document}
\begin{titlepage}
\maketitle
\end{titlepage}
\newpage
\begin{huge}
\textbf{Co-Authorship}
\end{huge} \leavevmode
\vspace{0.5in}\\
The following publications include a part of my thesis:

\begin{itemize}
  \item \textbf{Amine Barrak}, Ellis E. Eghan, Bram Adams. "On the Co-evolution of ML Pipelines and Source Code". IEEE International Conference on Software Analysis, Evolution and Reengineering (SANER2021). 
\end{itemize}

The following publication is not directly related to the material presented in this thesis, but were produced in parallel with the research performed for this thesis.

\begin{itemize}
  \item \textbf{Amine Barrak}, Ellis E. Eghan, Bram Adams, Foutse Khomh. "Why do Builds Fail? – A Conceptual Replication Study". Journal of Systems and Software (JSS2020). 
\end{itemize}
\newpage

\clearpage

\tableofcontents
\newpage

\listoffigures
\newpage

\listoftables
\newpage
\begin{abstract}

In the last few decades, companies are interested to adopt an online automated recruitment process in an international recruitment environment. The problem is that the recruitment of employees through the manual procedure is a time and money consuming process. The manual recruitment process could also possibly be erroneous in hiring incompetent individuals. As a result, processing a significant number of applications through conventional methods can lead to the recruitment of clumsy individuals. Different JD/Resume matching model architectures have been proposed and reveal a high accuracy level in selecting relevant candidates for the required job positions. 
However, the development of an automatic recruitment system is still one of the main challenges. The reason is that the development of a fully automated recruitment system is a difficult task and poses different challenges. For example, providing a detailed matching explanation for the targeted stakeholders (candidate recruiter, company who posted the job) is needed to ensure a transparent recommendation. 

There are several ontologies and knowledge bases that represent skills and competencies (e.g, ESCO, O*NET) that are used to identify the candidate and the required job skills for a matching purpose. Besides, modern pre-trained language models are fine-tuned for this context such as identifying lines where a specific feature was introduced. Typically, pre-trained language models use transfer-based machine learning models to be fine-tuned for a specific field. However, a combination of ontologies knowledge bases with modern language models is missing. In this proposal, our aim is to explore how modern language models (based on transformers) can be combined with knowledge bases and ontologies to enhance the JD/Resume matching process. 
Our system aims at using knowledge bases and features to support the explainability of the JD/Resume matching. Finally, given that multiple software components, datasets, ontology, and machine learning models will be explored, we aim at proposing a fair, explainable, and traceable architecture for a Resume/JD matching purpose.

As a first step, a systematic literature review is conducted to understand the available models of resume/ job matching architecture, the features used to address the matching, and the evaluation metrics used in the experiences.

Results of this thesis are targeted to make such e-recruitment become suitable for a fair JD/Resume matching; providing an explanation to the concerned stakeholders and keep a traceable, scalable JD/Resume recommendation system environment. The machine learning models' performance will be evaluated on a gold dataset provided by Airudi, using the normalized discounted cumulative gain according to the number of recommended candidates.

\textbf{Keywords:} Job Matching, Traceability, Explainability, Machine Learning.

\end{abstract}

\newpage
\section{Introduction}
\subsection{Context and Motivation}


Determining a suitable candidate for the job is not a simple task. The conventional recruitment process typically follows manual procedures. The manual recruitment process requires substantial sources such as trained recruiters in the human resource (HR) department, training expenses, etc. Moreover, these recruitment processes also require significant efforts and time to find relevant candidates for the required job positions. 
Therefore, filtering the most relevant candidates manually from a giant list of prospective candidates is troublesome.    

Several recent studies have been devoted to addressing the challenges related to the manual recruitment process. 
In the advertisement of job descriptions and recruitment processes, dealing with resumes in multiple languages is not easy. One of the most crucial challenges in multilingual job offers and resumes is finding the most relevant multilingual candidates through the manual recruitment process. For example, people speak multiple languages in countries such as Canada, India, and Belgium. Notably, in Canada, some people speak English, while others speak French in different cities (e.g., Montreal). Similarly, residents of Flanders communicate in Dutch. Nonetheless, Belgium has three official languages (Dutch, German, and French). Similarly, India has two official languages (English and Hindi). Hence, this implies that a larger pool of candidates in different languages seek job opportunities. 
Thus, an automatic recruiting system is required to help job seekers access the recruitment opportunities effectively and reduce the manual work in the recruitment process.

An eﬀective e-recruiting model frees companies from data overburden and advertisement cost, since it filter out incompetent candidates. The e-recruiting model can also help job seekers effectively access recruitment opportunities and reduce recruitment work. The key module for a unique e-recruiting model is the job matching framework that makes an eﬀort to draw in the jobless who are appropriate to the opportunities to be ﬁlled, where \textit{appropriate} means that a considered employer would be keen on perusing the retrieved resumes (curriculum vitae), while job seekers would have a fair chance to be hired. Finally, an automatic resume matching system can be significant in filtering relevant candidates during the recruitment process. Moreover, resume screening is a sensitive subject in biased decision making \ie{ethnic minority application} \cite{derous2020reducing}. Since machine learning models are trained using data, and if the data focuses on specific features, then machine learning models will make biased predictions that can have detrimental effects. Therefore, it is vital to ensure that the data is not biased and contains multifaceted classes. For example, training a model on people's resumes in a specific age range will create a biased model that may eliminate a qualified person.

The current job searching systems are unable to understand the semantic of various resumes and have not kept pace with the ongoing advancement in ML and natural language processing (NLP) methods. These solutions are commonly applied by manually extracted features/attributes and a set of rules with predefined weights on keywords that lead to an ineffective search experience for job-seeking candidates. Moreover, these techniques are not scalable. Moreover, some job seekers or company owners often keep fields empty in which information is required. For example, these fields can be job title, biography, etc. 

The data related to recruitment is usually handled by a relational database query system \cite{paoletti2016top}. An ideal framework would extract the exact features from applicant resumes for a job or several jobs possibly reasonable for an applicant. However, utilizing a relational database system for this job matching problem will run into two following significant barriers: (i) numerous text input fields are as free form or informal text by seekers instead of special keywords related to jobs. That implies that the desired output cannot be reliably matched; this is more of an information retrieval task, (ii) a number of fields are missing: applicants usually do not include all the fields in an online resume form. For instance, in the collection of a study, 90\% of resumes are missing the Summary field, and 23\% of the resumes are without the Resume Body field \cite{yi2007matching}.

The job recommendation systems require instantly to recommend accurate and precise jobs to the applicants and managers and regularly update the strategy of the system to maximize applicants' fulfillment. To accomplish personalization, applicants' explicit data, for instance, applicants’ jobs’ type, skills, experiences, age, gender, and salary package ought to be adequately utilized. Therefore, recommendation depends on explicit data that could bring risks of longer jobless duration or a large number of disappointed employment searchers \cite{reusens2017note}. That is one of the reasons that big companies like Microsoft recommend applicants to submit their implicit information (which is not explicitly present in a JD), for example, social networking sites (Facebook, LinkedIn, etc.), to consider applicants' online interactions. Implicit data comprises of all signs about applicants' interests that can be concluded by their online actions such as the sites they explored, the time they spent on a particular page, and the sites they bookmarked for returning to \cite{reusens2017note}. A job experience of a candidate had may contain implicit skills that were not mentioned, therefore, a semantic analysis of such experience can be understood in, similar context such as the intention may exist in JD \cite{gugnani2020implicit}.

There are several hurdles in modeling multilingual CV matching systems. From one perspective, a lack of resources and insufficient data to train machine learning algorithms, in particular, for a specific language, can lead a machine-learning algorithm to provide poor results. Therefore, the development of relevant datasets, especially annotated datasets can help to train a machine learning algorithm that can learn general hidden patterns in the datasets and obtain good performance. Such knowledge may be retrieved from structured public ontologies, which is a graph representation of semantic knowledge information (e.g, ESCO, O*NET). Annotated domain ontologies contain knowledge \ie{skills, education, universities} that can be refined with the additional dataset by conserving its internal associations' rules (same as, related to) \cite{chaudhary2018automated}.


Context-based transfer learning models \cite{vaswani2017attention}, such as BERT, XLNet, etc. have been very beneficial in producing state-of-the-art results in different NLP tasks, such as natural language understanding \cite{devlin2018bert}, language inference \cite{conneau2017supervised},  and machine translation \cite{vaswani2017attention}. Transfer Learning has also performed extraordinarily in the computer vision field where an essential step is to fine-tune the pre-trained models with ImageNet \cite{yosinski2014transferable, deng2009imagenet}. Some Simple Transformer models keep on advancing the field of NLP at a great pace, for example, DistilBERT, and  RoBERTa \cite{liu2019roberta}.

A language model system may identify correctly features in a Job/Resume, once it is fine-tuned on a large specific knowledge base.

Traceability and explainability are vital for transparency. Traceability is the ability to track every aspect of the process to improve product quality, operational efficiency, and the rise in safety awareness. In addition to this, traceability helps to review the product development flow. Traceability is essential to establish a communication connection and to promote collaboration with suppliers by implementing tracking systems. On the other hand, explainability aims to address how machine learning algorithms make a decision. Furthermore, explainability is an essential aspect of digital product development because it highlights the data insights, parameters, and decision point that machine learning algorithm used for decision-making and recommendation process. Consequently, Traceability and explainability are significant to minimize opaqueness.

\subsection{Objectives and Contributions}
This project aims to propose an effective e-recruiting tool to suggest the best candidates for the job postings. We propose to investigate the following objectives:
\begin{enumerate}
    \item study the State-of-the-art in the JD/Resume matching systems. 
    \item propose an e-recruiting architecture that considers JD/Resume matching by combining knowledge bases with a pre-trained transformer-based machine learning model such as BERT. 
    \item provide an explainable report to the stakeholders of the recommendations of the matching decision.
    \item adapt an existing traceable model to track the proposed matching and explainable architecture layers.
\end{enumerate}

This project will be accomplished by collaborating with a startup called \textit{"Airudi"} under a Mitacs internship program \footnote{\url{https://www.mitacs.ca/en/companies}}. Airudi aims to develop an e-recruiting tool that can recommend the best candidates to companies according to the job requirements. Moreover, Airudi is a third-party company that receives job offers from companies that require new people to fill various job positions. So, Airudi advertises the job offers and receives a list of prospective candidates interested in the job positions. Finally, Airudi is required to provide a list of the most appropriate candidates to the recruiters to conduct interview sessions. The e-recruiting system will recommend a list of resumes written in the same language required in the job description. For example, if a job description is written in French, then the e-recruiting system will only find the most relevant candidates having their resume written in French.



To achieve our goal, the following questions are designed to study a traceable, explainable, and fair JD/Resume recommendation system. 

\begin{itemize}
\item \textbf{RQ1: \RQOne}
In this research question, we plan to study the state-of-the-art in the JD/Resume matching. A systematic literature review will be conducted on works not earlier than in 2014 to cover the most recent developments concerning job description and resume matching.
\end{itemize}

After studying JD/Resume matching systems, our objectives will be based on two types of approaches/representations: 

\begin{enumerate}
\item \textit{The Ontologies and knowledge bases}: This language model uses a multi-relational graph that contains connected entities called nodes and relations called edges to create a structured representation (ESCO\cite{de2015esco}, DBpedia\footnote{http://mappings.dbpedia.org/index.php/Main\_Page}, WordNET\footnote{http://compling.hss.ntu.edu.sg/omw/}).

\item \textit{Transfer learning using language modeling}: These language models are first trained on a huge amount of text, known as pre-trained models such as BERT \cite{devlin2018bert}, MUSE \footnote{https://github.com/facebookresearch/MUSE}, and mBART \cite{liu2020multilingual}). The pre-trained models can learn the words, grammar, structure, and other linguistic features of a language. In addition to this, the pre-trained models can be fine-tuned on specific tasks such as classifying sentences in the resume or the job description if the sentences contain skills features \cite{9191408}.
\end{enumerate}

More specifically, we also intend to investigate the following research question:
\begin{itemize}
\item \textbf{RQ2: \RQTwo}

In this research question, our goal is to combine the multilingual knowledge provided by existing ontologies, i.e., ESCO, DBpedia, with fine-tuned modern pre-trained models to improve the identification of the multilingual features. Then, a matching process between the identified features in both JD/Resume will be adopted to rank the most appropriate candidates for the proposed job offer. 
Moreover, to verify if the proposed JD/Resume matching model is not biased in the token decisions will be considered.

\end{itemize}



Once a fair matching model is set up, a matching decision is made. Moreover, stakeholders need to have an explanation of the models taken decisions. We consider stakeholders as a job seeker, recruiter, or the company who posted the job. A detailed explanation of the concerned stakeholders is needed. Therefore, we formulate the following research question:

\begin{itemize}

\item \textbf{RQ3: \RQFour}

In this research question, we want to explore a way to improve the JD/Resume matching model designed previously to provide an explainable report to the concerned stakeholders. A report contains the list of the best-ranked candidates to \textbf{the recruiter} that contain information like (1) the selection criteria of the candidates and (2) a comparison between candidates to make a better decision during the interview day. A report to the \textbf{job seekers} indicates the possible reasons to rank taken decision (admitted/refused). A report for the \textbf{company who posted the job} will explain the recruiter evaluation criteria in choosing a person from the best-ranked candidature nominated for the job.

We will focus on the data cleaning to minimize bias and ensure the stakeholders' confidence in our explained reports.
\end{itemize}

In the previous \textit{RQ3}, a matching decision between resume and job description needs to be explained to the concerned stakeholders. During that process, there is a need to track the different stages concerning the evolution of the model's decisions by adapting existing traceability modules \cite{MLflow,dvc,atlas}. Therefore, we formulate our following research question: 

\begin{itemize}
\item \textbf{RQ4: \RQFive}

To answer this research question, a traceable module needs to be adapted for the current matching JD/Resume challenges described as follows: (1) Once a fair and explainable model is set up, multiple related submodels will be generated in the same pipeline (or workflow); (2) The most relevant features of resume and job description can be extracted using semantic models and/or deep learning methods can be adapted; (3) Experimental scenarios can be realized with a different set of values or hyperparameters to deploy the selected models; (4) To find the accurate models, the automation of these pipelines is essential and these previous steps must be repeated with a different set of parameters. By knowing that additional features such as the traceability tools may increase the system's complexity \cite{KARLSEN201635}, a case study on the co-evolution of ML pipelines and source code can be significant.

\end{itemize}

\section{Background}

\subsection{Basic Concepts}
We describe in the following subsections the main basic concepts that we will use during this proposal.
\subsubsection{Job Description (JD)}
A job description is a written description of what the person holding a particular job is expected to do, how they must do it, and the rationale for the required job procedures \cite{catano2009recruitment}.
A typical job description includes information about the company, contact details, job tasks, skills, and educational requirements, and desired personality. It may contain other details describing specific requirements for the job seekers' candidacy.

\begin{figure}[H]
\centering
\includegraphics[width=87mm,scale=0.2]{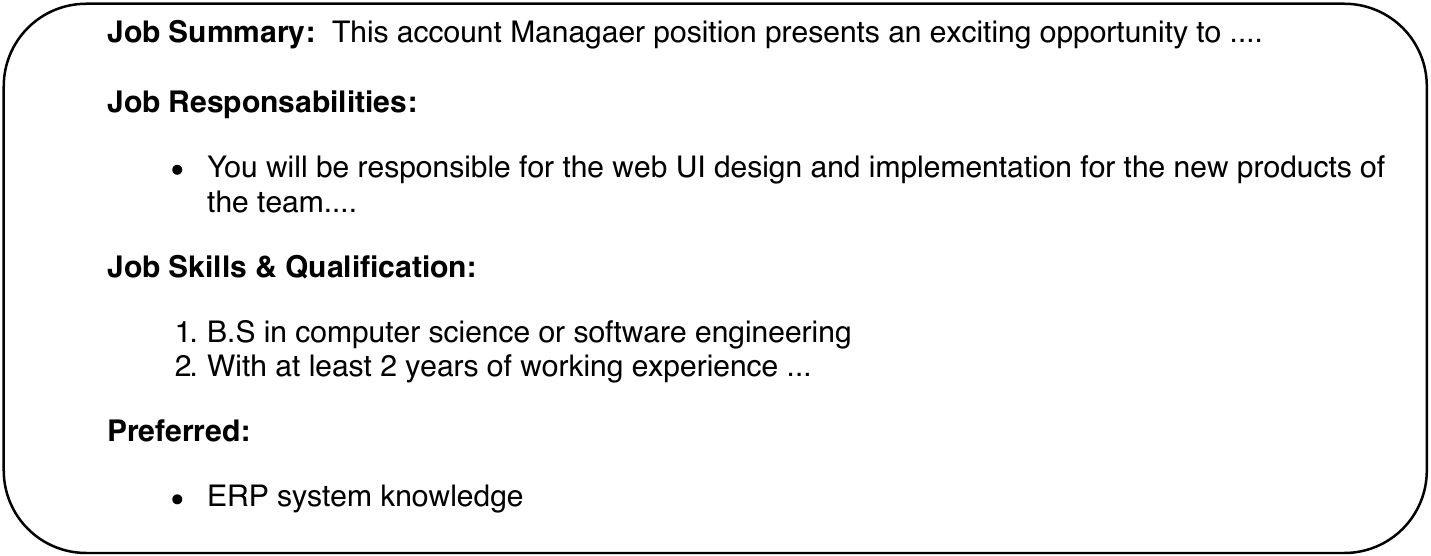}
\caption{Illustration of Job Description (JD)}
\label{JD}
\vspace{-5pt}
\end{figure}

\subsubsection{Candidate or Job Seeker}
A job seeker is someone who is looking for a job(s). He should present a resume that contains personal information, educational studies, skills acquired, his job experiences, languages mastered, etc. 

\subsubsection{Match a job seeker to a job description}
When a job seeker is looking for a specific job, the candidate will apply for the job and send his resume to the company that posted the job. Based on the job seeker's resume, and the job description details, a matching engine can use information parsed from the job description requirements and the list of resumes that applied to the job, such as, skills, education, degree of study, proficiency in languages, etc. Based on the similarity between a job description and a list of candidates, the matching engine will automatically recommend a list of the most similar resumes that meet the requirements. Finally, this automated process reduces the time to search for candidates and jobs using traditionally used listing providers and manual search techniques with the keyword.

\subsubsection{Data and ML pipeline traceability}

Modern ML applications require elaborate pipelines for data engineering, model building, and releasing \cite{amershi2019software}. 
Data engineers use a pipeline of tools to automate the collection, preprocessing, cleaning, and labeling of data. In contrast,  data scientists use a pipeline to extract the useful features from the data engineers’ data, execute machine learning scripts while experimenting with different sets of values for hyper-parameters, validate the resulting models, and then deploy and serve the selected models. Since these steps have to be repeated over and over whenever the data and/or model scripts or parameters change, in search of ever more accurate models, automation of these pipelines is essential.

Recently, a variety of data and model versioning tools have appeared to support data engineers and scientists \cite{atwal2020dataops}. Popular tools comprise DVC [4], MLFlow [5], Pachy-derm [6], ModelDB [7] and Quilt Data [8]. They typically combine the ability to specify data and/or model pipelines, with advanced versioning support for data/models, and the ability to define and manage model experiments.

One or many of these mentioned tools will be adapted to cover the traceability of the different layers of the proposed architecture.
\subsubsection{Biases in automated e-recruitment}

The biases in the decision of automated e-recruitment that can be linked to the trained machine learning models. A biased model could be trained on a specific type of people, \eg gender. The model in such a case will prefer a class of candidates compared to others.

\subsection{Information Retrieval Concepts}

\subsubsection*{Traditional word vector} 

\textit{Bag of Words or vector representation}. Bag of words (BoW) is a language model used to represent the presence or absence of a word. This language model provides a dictionary of words, but incapable of analyzing the relationships between words syntactically (structure) and semantically (meaning). 

\subsubsection*{TF-IDF}
The term frequency and inverse document frequency (tf-idf) is a weighting scheme used to assign 
a numerical statistic that is intended to reflect the importance of each word in the document. It is important to highlight that BoW model only creates vectors of word occurrences (counts). TF-IDF model, on the other hand, highlights what words are more important words and what words are less important in the dataset. BoW language model has such limitations such as this model does not take word ordering into account. Similarly, BoW model considers rare words less important. Therefore, to overcome these limitations, TF-IDF vectors can be vital. The Tf-idf is calculated as follows:

\begin{equation}
    W_{i,j}= tf_{i,j} * log(\frac{N}{df_{i}})
\end{equation}
Where:
\begin{itemize}
    \item $tf_{i,j}$ = Number of occurences of i in j
    \item $df_{i}$ = Number of documents containing i
    \item N = Total number of documents
\end{itemize}

\subsection{Evaluation Metrics}
The following section presents some basic common performance metrics used in literature experiments to evaluate their methodologies in the JD/Resume matching \eg performance of the model predicted classes of trained pairs of <JD, Resume>. Moreover, other classification metrics are considered according to the number of candidates that will be predicted \eg average precision.



\subsubsection{Performance Evaluation Metrics}
\label{evaluation}
In binary classification, a confusion matrix is commonly used to report performance metrics results. The Confusion Matrix (CM) is used in Table \ref{CM}\cite{hasanin2020investigating}.

\begin{itemize}
    \item True positive (TP) are positive instances correctly identified as positive.
    \item True negative (TN) are negative instances correctly identified as negative.
    \item False positive (FP), also known as Type I errors, are negative instances incorrectly
    identified as positive.
    \item False negative (FN), also known as Type II error, are positive instances incorrectly
    identified as negative.
\end{itemize}

\begin{table}[H]
\begin{tabular}{|l|l|l|}
\hline
                 & Labelled positive   & Labelled negative   \\ \hline
Positive prediction & True Positive (TP)  & False Positive (FP) \\ \hline
Negative prediction & False Negative (FN) & True Negative (TN)  \\ \hline
\end{tabular}
\caption{Confusion matrix for a binary classification}
\label{CM}
\end{table}

From the confusion matrix, we may calculate other performance metrics as shown in Table \ref{performance}. 

\begin{table}[H]
\begin{tabular}{l|l}
Performance metric                     & Formula \\\hline
Recall (R), True Positive Rate (T P R) &  $\frac{T P}{T P + F N}$       \\
True Negative Rate (T N R)             &  $\frac{T N}{T N + F P}$       \\
False Positive Rate (F P R)            &  $\frac{F P}{F P + T N}$     \\
Precision (P)                          &  $\frac{TP}{T P +FP}$       \\
Accuracy                               &  $\frac{T P +T N}{T P +T N+F P +F N}$  \\
F1-score                               &  $\frac{2*P*T P R}{P +T P R}$    \\

\end{tabular}
\caption{Common performance metrics using the confusion matrix.}
\label{performance}
\end{table}

Another metric Area Under the ROC curve (AUC) is used to avoid the usage of (TPR) and (FPR) independently. It is a plot of (TPR) versus (FPR) at different classification cut-offs. 
The Receiver Operating Characteristic (ROC) curves are usually used when there are roughly equal numbers of instances for each class, in other words, when the data is balanced \cite{saito2015precision}.

\subsubsection{Normalized Discounted Cumulative Gain}
The Discounted Cumulative Gain (DCG) is used in rankings with multiple grades of relevance, e.g., very relevant, relevant, irrelevant and very irrelevant \cite{cabrera2019ranking}.

The Normalized DCG (NDCG) is a performance metric that has seen increased adoption within the field of information retrieval \cite{lofi2015measuring}. It has been used in \cite{dehghan2020mining, balachander2018ontology, le2019towards, guo2016resumatcher, tran2017comparison, grover2017latency, reusens2018evaluating}.

\begin{equation}
    DCG_{n}=\sum_{i=1}^{n}\frac{2^{rel_{i}}-1}{log_{2}(i+1)}
\end{equation}
\begin{equation}
    NDCG_{n}=\frac{DCG_{n}}{IDCG_{n}}
\end{equation}
\begin{equation}
    IDCG_{n}=\sum_{i=1}^{|rel|}\frac{2^{rel_{i}}-1}{log_{2}(i+1)}
\end{equation}

Where:
\begin{itemize}
    \item Value of rel i is 1 if the item at i in the ranked list is
correct recommendation, otherwise rel i is 0.
    \item n: length of the returned list.
    \item DCG n : is DCG value of the TopN.
    \item IDCG n : is ideal DCG value of the TopN.
    \item |rel| is the size of the jobs.
\end{itemize}

\subsubsection{Average Precision}
Average Precision (AP) measurement is used to rank two grades of relevance (relevant and irrelevant). These measurements determine how accurate the recommendation system ranks candidates' applications and the selected candidates. These methods generate a score according to the rank of actually recommended applications on the top-k recommendation list.

\begin{equation}
    AP = \frac{\sum_{i=1}^{n}(P(i)*rel(i))}{\#releventitems}
\end{equation}
\\
Where:
\begin{itemize}
    \item n: the number of recommended jobs for a user.
    \item rel(i) is 1 if the item at position i in the ranked list is
correct recommendation, otherwise it is 0.
    \item P(i) is precision of top i.
\end{itemize}

In addition to this, some studies also used a mean average precision to evaluate the performance of machine learning models \cite{cabrera2019ranking, luo2019resumegan, ozcan2016applying, tran2017comparison}.

\subsubsection{MRR (Mean Reciprocal Rank)}
Reciprocal rank (RR) is a measure that takes into account the first position of the relevant ranked resume list. MRR is the mean of all jobs’ RR values. This measure was considered in \cite{dehghan2020mining, le2019towards, tran2017comparison, fernandez2019cv, gugnani2020implicit}.
\begin{equation}
    MRR=\frac{1}{\left | U \right |}\sum_{i=1}^{U}\frac{1}{Rank_{i}}
\end{equation}
\\
Where:
\begin{itemize}
    \item $\left | U \right |$: The number of jobs have recommendation users
    \item $Rank_{i}$ : The first relevant position in recommended users
\end{itemize}


\section{Systematic Literature Review}
\label{slr}
We designed a Systematic Literature Review (SLR) to cover the existing research done in the domain of JD/Resume matching. In the beginning, we start by describing the methodology we followed in our SLR in section\ref{method_slr}. We present the methods used for artificial intelligence (IA) explainability in section  \ref{explainable} the different features used in resume/ job description in section\ref{features} and the system knowledge representation in section\ref{semantic}. Furthermore, in section \ref{ml}, we provide an overview of machine learning-based recommendation systems. Recommendation models are presented in section\ref{lingual}. Eventually, we talk about Machine learning traceability systems\ref{traceability}. 

\subsection{Methodology}
\label{method_slr}
A systematic literature review is considered an effective research methodology \cite{Kitchenham07guidelinesfor} to identify and discover new facts about a research area and to publish primary results to investigate research questions \cite{Staples:2007,Kitchenham07guidelinesfor}.

 This SLR is used to achieve the following five objectives:
\begin{itemize}
    \item Understand the JD/Resume matching. 
    \item Identify the features used in the literature to make the matching.
    \item Categorise the different methodologies used to match JD/Resume.
    \item Find different metrics used to evaluate the matching process.
    \item Investigate the methods used to cover multilingual JD/Resume
\end{itemize}

The relation of this SLR with the thesis goal is to create a catalog of the most used methodologies of multilingual JD/Resume matching and identify the gaps inside.

To the best of our knowledge, in the literature, there is no systematic literature review on matching JD/Resume published for the period we covered between 2014 and 2021.

\subsubsection*{SLR Planning}
We performed an SLR covering matching between resumes and job descriptions in human resources published from 2014 to 2021. Instead of applying a manual search, we perform an automated search using Engineering Village \footnote{\url{https://www.engineeringvillage.com/search/quick.url}} to search for papers related to the matching of resumes and job description. Engineering Village is an information discovery platform that is connected to several trusted engineering electronic libraries. Specialized in engineering, it offers many options to refine the search queries, excludes and inclusion criteria, and provides the flexibility for the choice of period, language, venues, and authors.

This platform gives users also the ability to search for all recognized journals, conference, and workshop proceedings together with the same search query \cite{Sharafi:2015}.

Engineering village includes three data banks \texttt{Compendex}, \texttt{Inspec}, and \texttt{Knovel}. For our study, we will focus on one data bank, \texttt{Compendex}, to avoid duplicated papers.

According to our goal which is to study the JD/Resume matching system in the literature, we assume that the main keywords to make the search query are: \textbf{Resume}, \textbf{Job Description} and \textbf{matching}. We have used keywords, their synonyms, and stems to make our search query. Synonyms and truncations are needed to ensure a complete collection of papers.
\begin{enumerate}
\item Resume: resume*, cv, candidate, employee*, job seeker
\item Job: job*, "human resource", recruiter
\item Match: recruitment* OR recommendation* OR hire OR hiring OR match*
\end{enumerate}

Using these keywords, we have combined them with logical operators (AND, OR). The final search query is:
\begin{center}
\textit{
((resume* OR cv OR candidate OR employee* OR "job seeker") AND (job* OR "human resource" OR recruiter ) AND (recruitment* OR recommendation* OR hire OR hiring OR match*) AND ({ca} OR {ja}) WN DT) OR ("person-job" AND fit*)
}
\end{center}
 
\texttt{"(\{ca\} OR \{ja\}) WN DT"} is an attribute used to allow the server of Compendex to limit results to only documents of type conference articles or journal articles.

Please note that we validated the query on a set of papers that we knew already relevant. 

\subsection*{SLR Execution}
The SLR execution phase was carried out in two steps and executed in August 2020. The first one was dedicated to the execution of the search query on the Engineering village platform. 

The query returned \textbf{752} papers as primary results. Browsing the research articles rapidly, we found that unrelated terms needed to be excluded \ie sentiment, behavior, work turnover, sales, work stress, jobseeker satisfaction, crime, appearance, social network.

We proceeded to add exclusion criteria to exclude papers out of scope. We used the check-box feature offered by Compendex to make the exclusion (\textit{turnover OR satisfact*, stress*, emotion*, appear*, crime*, sale*, advertis*, behavior*, "social network", sentiment*}). The search was limited also to \texttt{English language}. Only conference and journal papers were retained. 

We analyzed all the research papers and verified their relevance to our study, in case we found a relevant one, we included it in our paper catalog (9 papers were manually added).

Finally, we collected \textbf{514} articles with the JD/Resume matching. 

The second step was dedicated to the manual analysis of the collected articles. We performed three rounds:
\begin{itemize}
\item The first round was reading the abstract, introduction, and conclusion of the \textbf{514} papers and eliminating irrelevant or research articles having page length of fewer than 4 pages (3 papers). Our data included \textbf{85} conference and journal papers after applying the first round.

\item The second round was dedicated to the snowball search technique\footnote{to mitigate the fact that we considered only compendex databse}. We used it to run through all the paper references and extract if any, articles that were missed or that the search method was unable to identify. Our data became \textbf{109} papers after the snowball round.

\item The third round was committed to particularly focus on the 109 papers. A complete reading of articles was performed to extract:
\begin{itemize}
\item The features used to realize the matching.
\item The established methods to extract features.
\item The matching process.
\item The evaluation metrics.
\end{itemize}
\end{itemize}

We present the SLR results and the related work of this thesis in the following sections, organize as follows. First, we discuss the state-of-the-art of explainable AI in section \ref{explainable}. Next, we describe the different features used to deal with JD/Resume matching in section \ref{features}. In section \ref{semantic}, we present the knowledge base models and the machine learning architectures in section \ref{ml}. In Section \ref{lingual}, we provide a comparison between multilingual matching models. Then, we present the possible biases in machine learning algorithms in section \ref{biases_ml}. Section \ref{traceability}presents the data and machine learning traceability models.



\subsection{Explainable Model Architectures}
\label{explainable}
JD/Resume matching models have complex architectures. The matching or non-matching decision of these models is difficult to understand. The different stakeholders (\textbf{recruiter}, \textbf{job poster}, \textbf{job seeker}) related to a JD/Resume matching need a personalised addressed explanation. For example, the company who opened the job vacancy should receive the reasons that make a list of candidates more suitable to their job description from the company who posted the job. Therefore,  any detail (features) in the resume and job description should be interpreted.

Explainability and interpretability are often used interchangeably to understand the reasons how artificial intelligence (IA) models made decisions in matching JD/Resume matching. Therefore  Explainability and interpretability are vital to understanding the model's decision-making process.  Moreover, the \textbf{interpretability} is used to understand a cause and effect relationship within a system. For example, understanding what features are more important and helpful in matching model decision-making process. \textbf{Explainability}, on the other hand, is used to study the internal mechanics of a machine or deep learning system so that the model matching decision can be explained in human terms \cite{MachineL86}.

Previous studies \cite{qin2020enhanced, qin2018enhancing} used the model's interpretability to highlight the most important features given by the attention model in a resume or job post matching. For example,  Le \al~\cite{le2019towards} reported that the interpretability could be summarised using an intention rate model of the job seeker and the employer. Likewise, another study by Jiang \al~\cite{jiang2020learning} revealed that the features extracted from resumes as semantics entities are helpful in interpreting the matching result. Finally, all the machine learning models are trained using data, therefore insights about data extraction and collection process can be crucial in JD/Resume matching process. 

The explanation data extraction process should be explicitly explained to keep the matching traceability during the whole process. However, deep learning models are typically difficult to interpret due to complex internal transformations and considered as a black box \cite{lipton2018mythos}. Most importantly,  some initiatives have taken to overcome this issue \cite{haffar2020explaining,bau2017network}.

According to the best of our knowledge, no JD/Resume matching architecture has been proposed as an explainable model yet. However, different studies have been conducted to highlight the importance of explainability in machine learning decision models. For example, a recent study by Danilevsky \al~\cite{danilevsky2020survey} realized a survey on the explainability of IA for natural language processing and reported the operations that enable explainability. These operations are: (1) \textit{Layer-wise relevance propagation} \cite{poerner2018evaluating}, (2) \textit{input perturbations} (3) \textit{Attention base models} feature importance \cite{luo2018beyond}, (4) \textit{LSTM} and feature importance explainability \cite{ghaeini2018interpreting}, and (5) \textit{Explainability-aware architecture design} \cite{liu2018interpretation}. Particularly, Layer-wise relevance propagation is used to enable feature
importance explainability. Similarly, input perturbations usually used for a linear model LIME and  Attention based models are used to highlight important features. Similarly, another study \cite{ghaeini2018interpreting} presented LSTM and feature importance explainability, and Explainability-aware architecture design \cite{liu2018interpretation}.

Le \al~\cite{le2019towards} tried to overcome the interpretability problem by comparing the intention of the job seeker and employers. However, this is still far from having good reasons that explain the matching decision reasons.

\subsection{Job Description and resume features used for matching} 
\label{features}



In the provided literature, features are divided based on their usage in candidate resumes and job descriptions from the employers. These are distributed as education level, skills, personal information, job history, experience, and job industry information provided by the candidates. On the other hand, the required information in the job description consists of the same information as for resumes and salary packages offered and jobs to perform in the specific industry.

Zhang and Vucetic \cite{zhang2016sampling} conducted a case study on Linkedin with graduated students from the same university where they found that features considered to be important in the recommendation of resumes to the job offer were not used \ie year of graduation, gender, and grade point average. This depicts that there is a gap of research to be done with respect to the grades and gender of the candidates.

\subsubsection{Resume Features}
\textbf{Education} The education section involves education level, specialization in the relevant field, awards or achievements, and research publications. These features show the candidates' educational backgrounds. 

While conducting the resume analysis, education level or qualification information has been considered vital because of its role in matching with a suitable job. Some researchers also included academic awards and achievements as features in algorithms' design \cite{ramannavar2018proposed, gupta2019comprehensive, ramannavar2018proposed, nimbekar2019automated, javed2015carotene, kethavarapu2016concept, fernandez2019cv}.  
Thus, approximately every research study has included it as a resume feature. However,  some studies primarily focused on skill analysis \cite{qodad2019adaptive, corde2016bird, nigam2019job, balachander2018ontology, singh2017propensity, gatteschi2016learning}, and job descriptions' features \cite{liu2016companydepot, kusnawi2019decision, liu2016employer, qin2018enhancing, reusens2018evaluating, nigam2019job, papoutsoglou2017mining, ahuja2017similarity, namahoot2017standard}. Other educational features include research paper publications and certification in the specific fields candidates are graduated in \cite{liu2017hierarchical, zhang2014research, sandanayake2018automated, nimbekar2019automated, qin2018enhancing, jainoptimizing, celik2016towards}.

Multiple studies have collected datasets from various fields such as IT \cite{chanavaltada2015improvement, martinez2019novel, truicua2019innovating, valverde2018job, espenakk2019lazy, balachander2018ontology}, programming languages \cite{qin2020enhanced, kethavarapu2016concept, fernandez2019cv, kusnawi2019decision, dehghan2020mining, walek2016proposal, zaman2018staff, paoletti2015extending}, software engineering \cite{zaroor2017hybrid, maree2019analysis, kmail2015automatic, ozcan2016applying}, Human Resources \cite{susanto2019employee}, Economics \cite{gatteschi2016learning}, Business \cite{chanavaltada2015improvement}, and computer sciences \cite{gupta2014applying}. In addition to this, other studies are based on available datasets from various recruitment sites (indeed, monster, glassdoor, amrood, careerbuilder, BOSS Zhipin and jobstreet) \cite{gupta2019comprehensive, guo2014analysis, kmail2015automatic, kmail2015matchingsem, guo2016resumatcher, bafna2019task, chen2018tree}, social media platforms (LinkedIn and Facebook) \cite{gupta2019comprehensive, almalis2014content, ramannavar2018proposed, grover2017latency, zaman2018staff}, government recruitment departments \cite{fernandez2019cv, liu2016employer, reusens2018evaluating, mrsic2020interactive} and university career centers \cite{gupta2019comprehensive, nguyen2016adaptive, liu2017hierarchical, yi2016job, liu2016employer, lee2020industrial, jacobsen2019s, jainoptimizing}. The datasets collected from universities are based upon the students' qualifications only \cite{gupta2019comprehensive, nguyen2016adaptive, liu2016employer, lee2020industrial, jacobsen2019s, jainoptimizing}.

\textbf{Acquired Skills} Skills are the natural or learned talents and the expertise developed by the candidates to perform a task or a job. There are several key types of skills: soft skills, hard skills, domain-general, and domain-specific skills. However, incorporating skills into resumes is not as simple as it sounds. There are different categories of skills to understand, for instance. Moreover, it's essential to select the right skills and to include them in resumes.

The second most important features while conducting the resume analysis are related to the skills obtained in a specific field. Likewise, technical proficiency while working in a specific job position, years of experience, and resume holders language proficiency. Some studies only used university datasets, however, the details such as the students have no relevant practical experience in their fields are missing \cite{gupta2019comprehensive, nguyen2016adaptive, liu2017hierarchical, yi2016job, lee2020industrial, jacobsen2019s, jainoptimizing}. A feature of the actual position is added by various algorithms to enhance the workability of job matching \cite{ramannavar2018proposed, duan2019resume, sandanayake2018automated, almalis2014content, tran2017comparison, maheshwary2018matching}. Finally, Some frameworks are presented to define language as a resume feature because some jobs require native or foreign-language speakers. Thus this can play a positive role in job matching and recommendation \cite{almalis2014content, yi2016job, roy2020machine, bal2016matching, guo2014analysis, kmail2015automatic, javed2015carotene, guohao2019competency, kethavarapu2016concept, fernandez2019cv}.

\textbf{Personal Features} In job recommendation and matching systems, researchers consider unique features in the resume to locate the relevant jobs depending upon the age, language, location, nationality gender, driving license, marital and military status. These features directly impact the job description requirements, and that is why considered important to be added. However, some studies used candidates' personal details without adding unique features \cite{liu2017hierarchical, almalis2015fodra, truicua2019innovating, grover2017latency, espenakk2019lazy, papoutsoglou2017mining, balachander2018ontology, zhu2018person, amin2019web, charleer2019supporting}.

The current location feature is required when the job is location specific, or the recruitment companies want to consider a candidate from a specific area \cite{tran2017comparison, gupta2019comprehensive, almalis2014content, nguyen2016adaptive, pahari2016framework, kmail2015automatic, qin2020enhanced, luo2015macau, xu2018matching, mohamed2018smart, zaman2018staff, gatteschi2016learning, chaudhari2015design}. In addition,  the age of the resume holders is considered as the next personal feature and the jobs are filtered based on  the age requirements by the job matching algorithms set by the recruiters \cite{gupta2019comprehensive, pahari2016framework, roy2020machine, fatma2020canonicalizing, he2019career, javed2015carotene, mrsic2020interactive, dong2017job, chaudhari2015design}. Studies have also considered gender information to filter gender-specific jobs and to make it easy for matching \cite{gupta2019comprehensive, roy2020machine, qin2020enhanced, xxxxneural, chanavaltada2015improvement, gupta2014applying, ozcan2016applying, ramannavar2018proposed, zhang2014research, wang2015resume, espenakk2019lazy, namahoot2017standard, liu2019tripartite, maheshwary2018matching}. Marital status feature of the candidates is also added to the personal feature library by some researchers \cite{gupta2014applying, ozcan2016applying, sandanayake2018automated, Leah_vectorisation, rodriguez2019feature, wenxing2015ihr+, guo2016resumatcher}. The next personal resume feature of applying candidate is a nationality, and it holds the same importance as location feature as it helps in addressing the workplace location and requires nationality to avoid any travel sanctions \cite{pahari2016framework, zaroor2017hybrid, bal2016matching, martinez2019novel, wang2015resume, chala2017knowledge, grover2017latency, espenakk2019lazy, luo2015macau, xu2018matching, dehghan2020mining, qiao2019mlca, celik2016towards, chen2018tree, maheshwary2018matching}. Only one research framework has included military status to the personal features library \cite{ozcan2016applying} and culture \cite{pessach2020employees}.

\textbf{Features Linked to Jobs} Resume features linked to candidates' job history, current position, salary scale, actual pay, and industry of the job are essential as they directly match the job requirements mentioned in the job description of respective fields. Actual pay \cite{roy2020machine, deng2018improved, gupta2014applying, ozcan2016applying, duan2019resume, chaudhary2018automated, chen2017hybrid, dong2017job} and salary scale \cite{tran2017comparison,  guo2014analysis, deng2018improved, chaudhary2018automated} are the resume feature to match the pay package offered by the company and thus considered by many researchers. The industry of the jobs of candidates is from an important feature to be considered to align with the technical job description features, and this is the reason nearly all the research studies include it in their matching algorithms \cite{tran2017comparison,  almalis2014content,zaroor2017hybrid, zaman2018staff, namahoot2017standard, gatteschi2016learning, paoletti2015extending, cernian2017boosting, maheshwary2018matching, chaudhari2015design}. Furthermore, some researchers used information about jobs demand in industry to better understand candidate's interests \cite{zaroor2017hybrid, roy2020machine, kmail2015automatic, chen2017hybrid, mrsic2020interactive, espenakk2019lazy, luo2015macau, xu2018matching, kmail2015matchingsem, walek2016proposal, maheshwary2018matching}. The candidate's experience is considered by taking two things into account: (i) his previous employment experience ( history in different companies)  \cite{he2019career, shalaby2016entity, zaroor2017jrc}, and (2) the number of jobs he applied in the past \cite{qodad2019adaptive, he2019career, borisyuk2016casmos, almalis2015fodra, nigam2019job, zaroor2017jrc, chen2018tree} have been taken as features by the researchers. From the employment point of view, employment preferences \cite{ozcan2016applying} and employee turnover \cite{pessach2020employees} are added as resume features.

\subsubsection{Job Description Features}

There is ample detail in the job description to identify major roles and important tasks as they occur today. They are not dependent on any particular qualities of an incumbent (such as experience, expertise, ability, efficiency, commitment, loyalty, years of service, or degree) \cite{yi2007matching}. They provide the details required to identify the job, not the employee.

\textbf{Personal Requirements} The job descriptions issued by recruitment agencies or companies possess a certain format which is based on the primary and secondary level of important information. As mentioned earlier, some companies are more intended towards getting technical and qualification information rather than personal details \cite{liu2017hierarchical, ramannavar2018proposed, lee2018artificial, chaudhary2018automated, reusens2018evaluating, almalis2015fodra, amin2019web, gatteschi2016learning, paoletti2015extending, charleer2019supporting}. Depending upon the vacancy available and suitable gender quota, gender information is considered to be important in job description analysis by studies undertaken in existing literature. \cite{ramannavar2018proposed, zhang2014research, wang2015resume, kethavarapu2016concept, fernandez2019cv, chanda2019developing, manad2018enhancing, shalaby2016entity, wenxing2015ihr+}. The required age for the suitable job is also significant to find a specific job \cite{gupta2019comprehensive, almalis2014content, pahari2016framework, liu2017hierarchical, roy2020machine, bal2016matching, chanavaltada2015improvement,  maheshwary2018matching, chaudhari2015design}. Some of the studies have also included civil status \cite{xxxxneural, zhang2014research, wang2015resume, rodriguez2019feature, chen2018tree, chaudhari2015design}, military status \cite{ozcan2016applying} and needed ability \cite{chaudhary2018automated, wenxing2015ihr+, singh2017propensity, kino2017text} as personal requirements features. Location of the job placement should be known for the candidate thus it is added frequently by the researchers \cite{pahari2016framework, zaroor2017hybrid, bal2016matching, maree2019analysis, guo2014analysis, susanto2019employee, pessach2020employees, maheshwary2018matching}.

\textbf{Educational Requirements} This section lists the required level of job knowledge (such as education, experience, knowledge, skills, and abilities) required to do the job. This section focuses on the “minimum” level of qualifications for an individual to be productive and successful in this role. In a job description, it is essential to identify the educational qualifications that an employee must possess to satisfactorily perform the job duties and responsibilities. \cite{Writinga29} Thus, the educational qualifications must be stated well in terms of areas of study and/or type of degree or concentration that would provide the knowledge required for entry into this position.

Educational requirements features such as degree names and grades have a primary significance in job recommendation systems and all studies have added this feature in job description analytic algorithms except a few that are more into technical skills \cite{qodad2019adaptive, corde2016bird, nigam2019job, balachander2018ontology, singh2017propensity, gatteschi2016learning}. The academic awards, i.e. scholarships and awards, are also considered for the distinctive recruitment of employees \cite{gupta2019comprehensive, liu2017hierarchical, chanavaltada2015improvement, ramannavar2018proposed, wang2015resume, nimbekar2019automated, qiao2019mlca}.

\textbf{Offered Position} The purpose of job descriptions is to make candidates understand the nature of their responsibilities depending upon their skills, ability and qualification. The job description must offer a suitable position for the candidates considering these requirements. Thus, various studies have distributed this feature into sub-categories for a better match result and improved the algorithm's performance \cite{Understa75}. All studies involving job description analysis include offered positions and industry types for which jobs are available \cite{tran2017comparison, gupta2019comprehensive, almalis2014content, nguyen2016adaptive, liu2017hierarchical, zaroor2017hybrid, yi2016job, nguyen2018linguistic, roy2020machine, guo2014analysis, kmail2015automatic, xxxxneural, chanavaltada2015improvement, ozcan2016applying, ramannavar2018proposed, zhang2014research, chou2019resume, wang2015resume, lee2018artificial, mehta2019service, chaudhary2018automated, nimbekar2019automated, palshikar2018automatic, Leah_vectorisation, malherbe2016bridge, fatma2020canonicalizing, patel2017capar, kethavarapu2016concept, fernandez2019cv, kusnawi2019decision, chanda2019developing, bian2019domain, susanto2019employee, manad2018enhancing, shalaby2016entity, malherbe2014field, almalis2015fodra, ep2019framework, jacobsen2019s, qiao2019mlca, walek2016proposal, mohamed2018smart, bafna2019task, kino2017text, maheshwary2018matching}. The offered salary is mentioned in job descriptions depending on the experience and skills a candidate brings to the position. \cite{nguyen2016adaptive,liu2017hierarchical, yi2016job, roy2020machine, guo2014analysis, kmail2015automatic, martinez2019novel, gupta2014applying, wang2015resume, mehta2019service, chaudhary2018automated, malherbe2016bridge, fatma2020canonicalizing, kethavarapu2016concept, fernandez2019cv, chanda2019developing, manad2018enhancing, shalaby2016entity, gugnani2020implicit, kmail2015matchingsem, dehghan2020mining, qiao2019mlca, amin2019web, maheshwary2018matching}. Depending upon the seniority of job and responsibilities, years of experience define the candidates' suitability, and this is why all the studies have included it as job feature except the ones that are considering the university datasets or fresh graduates. \cite{gupta2019comprehensive, nguyen2016adaptive, liu2017hierarchical, yi2016job, nguyen2018linguistic, zhang2014research, lee2018artificial, patel2017capar, guohao2019competency, chanda2019developing, liu2016employer, lee2020industrial, jacobsen2019s, jainoptimizing}. 

Companies have certain workplaces for their employees, such as to work in a team or individually. It is important to highlight that some studies considered the candidate working experience in a team or individually as a feature, called as teamwork skills\cite{nguyen2018linguistic, martinez2019novel, chaudhary2018automated, kethavarapu2016concept, kusnawi2019decision, qin2018enhancing, shalaby2016entity, almalis2015fodra, walek2016proposal, luo2019resumegan, cernian2017boosting} and work length \cite{almalis2014content, yi2016job, nguyen2018linguistic, guo2014analysis, kmail2015automatic, sandanayake2018automated, wenxing2015ihr+, valverde2018job} as requirements for the candidates.

\textbf{Technical Job Requirements}
A list of the technical roles and obligations allocated to the job is given in this section; the basic tasks are also referred to job requirements. The job requires sufficient knowledge of the subject area to address both unique and normal work challenges, to be able to comment on technological concerns, and to act as a guide for those within the organization on the subject. \cite{Writinga29} Thus, it is important to list particular abilities and/or skills needed for the performance of the candidate in this position, including the designation of any required licenses. Analytical, budget exposure, internal or external contact, machine, innovative thinking, customer service, decision-making, variety, critical thinking, multi-tasking, collaboration, problem-solving, project management, oversight, coordination, are some considerations:

In job description analysis, the technical required information such as technical information \cite{guo2014analysis, kmail2015automatic, kusnawi2019decision, bian2019domain, susanto2019employee, manad2018enhancing, chaudhari2015design}, technical categories \cite{zaroor2017hybrid, lee2018artificial, mehta2019service, ramannavar2018proposed, valverde2018job, walek2016proposal, chaudhari2015design}, and specific field experience \cite{zaroor2017hybrid, chou2019resume, lee2018artificial, mehta2019service, bian2019domain, susanto2019employee, manad2018enhancing, shalaby2016entity, chaudhari2015design} are necessary. Thus, these features are found to be essential for job resume matching algorithms. 

Furthermore, all these jobs and personal features are divided among certain matching and recommendation frameworks to differentiate among job recommendation, matching, content-based analysis, and resume analytics.
One to one job description and resume matching are used in the majority of the studies \cite{roy2020machine, bal2016matching, martinez2019novel, ramannavar2018proposed, zhang2014research, chou2019resume, wang2015resume, susanto2019employee, pessach2020employees, mrsic2020interactive, paoletti2015extending}. Apart from developing a matching model of both resume and job description matching, some researchers are more interested in only one of these, i.e., job recommendation. \cite{pahari2016framework, qodad2019adaptive, guo2014analysis, qin2020enhanced, gupta2014applying, mehta2019service}. These recommendation systems adopt certain algorithms by combining position description and resume information. These algorithms are content base analysis \cite{gupta2019comprehensive, nguyen2018linguistic, patel2017capar, shalaby2016entity}, ontology based \cite{kethavarapu2016concept, balachander2018ontology, cabrera2019ranking} and text base classification \cite{patel2017capar, borisyuk2016casmos, maheshwary2018matching}.






 
    
    





\subsection{Semantic Representation}
\label{semantic}
Semantic methods are useful to identify linked item ideas, since an idea can be described in multiple textual ways, relying on implicit knowledge of how different terms relate. This information can be encoded in taxonomies, where relations between different terms are mapped, which then can be used during job matching.

\subsubsection{Similarity Measures}
\label{similarity}
There is some work designed to make the matching based on the text similarity between the candidate resumes and the job description \cite{duan2019resume,neculoiu2016learning}. This method is based on transforming the list of features (\ie education, skills, years of experience, etc) extracted from the resumes and job descriptions into vectors. A popular measure in data science is the cosine similarity used to compute the angle difference between two vectors. The measure will equal 1 when the vectors are parallel (they point in the same direction) and 0 when the vectors are orthogonal. Vectors that point in the same direction are more similar than vectors that are orthogonal \cite{widdows2003orthogonal}.

\subsubsection*{Cosine Similarity}
Wenxing \al~\cite{wenxing2015ihr} proposed a mobile reciprocal job recommender based on computing the cosine similarity between feature vectors of the job seekers and the recruiters. Duan \al~\cite{duan2019resume} used the vector space model (VSM) to cluster the resumes based on their similarity to reduce the number of matching the job resume to each position by addressing only the match between clusters and job description. Rodrigues \al~\cite{rodriguez2019feature} classified candidates by feature similarity \ie work experience, education, etc. In contrast, Gubta and Garg \cite{gupta2014applying} proposed a personalized recommendation to the candidate according to his profile \ie preferring the company that has the same current location \cite{yu2011reciprocal}. Kenthapadi \al~\cite{kenthapadi2017personalized} discussed the personalized job recommendation strategy at LinkedIn where the job seekers receive personalized job postings based on the context data present in their profiles, activities and similar members. However, Nigam \al~\cite{nigam2019job} demonstrated that if some candidate applies for similar jobs according to their interests, this will be a subject of candidates motivation. For example, if some candidates applied for a job, the same candidates can also be interested in applying for other similar jobs.

\subsubsection*{Jaro Winkler distance}
Jaro Winkler distance \cite{winkler2006overview} is a measure of similarity between two strings, the higher the Jaro distance for two strings is, the more similar the strings are. Maree \al~\cite{Maree_2018} used this technique to compare the sense of a term used in resumes and job descriptions if it has a close distance based on the term surroundings words. Çelik~\cite{celik2016towards} measured the similarity between two terms to eliminate mi-spelling errors from resumes and jobs description in the parsing process.

\subsubsection*{LSI \& LDA}
The clustering of text and the calculation of similarity should be calculated on the basis of the text model. The commonly used models are latent semantic index (LSI) and Latent Dirichlet allocation (LDA).

Latent semantic indexing (LSI) is used to reduce the dimension for classification. The idea is that words will occur in similar pieces of text if they have similar meanings. It is an indexing and retrieval method that uses a mathematical technique called singular value decomposition (SVD) to identify patterns in the relationships between the terms and concepts contained in an unstructured collection of text \cite{deerwester1988improving}.

On the other hand, Latent Dirichlet allocation (LDA) has been used to identify the main topic (meaning) of a text. LDA works by creating a normal distribution of words by randomly choosing topics and then checks for the probability of the word to belong to a topic regarding all the documents \cite{jainoptimizing}. the highest score is chosen as the final topic. This method has been used in several works to extract the topic distribution from jobs or resumes \cite{qin2020enhanced, jacobsen2019s, jainoptimizing, dehghan2020mining}.

\subsubsection{Ontologies and knowledge bases}

We found in the literature of JD/Resume matching several approaches that improve or use the knowledge of existing ontologies or taxonomies to extract the list of skills in JD/Resume. For example, the ontology Occupational Information Network (O*NET)\footnote{https://www.onetcenter.org/} database in the USA, the multilingual European Dictionary of Skills and Competencies (DISCO)\footnote{http://disco-tools.eu/}, the 'European Skills, Competences, Qualifications and Occupations' (ESCO)\footnote{https://ec.europa.eu/esco/portal/home} have been extended or served as a base model to create a new ontology/taxonomy. The ontologies are made in a way that can be updated at any time and adapted to the dynamics of the labor market \cite{rentzschskills}.

More ontologies were used to extract semantics from the parsed JD/Resume, such as WordNet \cite{miller1998wordnet} which is a lexical resource of different domains that contains synonyms and hyponym relations between words. YAGO \cite{suchanek2007yago} is a crowdsourced platform containing structured and relational information extracted from Wikipedia and other sources in multiple languages. Similarly, DBpedia \cite{lehmann2015dbpedia} ontologies in different domains have been created based on the most commonly used infoboxes within Wikipedia.

\subsubsection*{Taxonomy}

In natural language processing, a taxonomy provides machines ordered representations and hierarchical relationships among concepts and the words employed to describe those concepts. For example, a basic NLP taxonomy would have concepts such as machine learning, which is a subset of AI, and deep learning, which is a subset of machine learning. In other words, a taxonomy is a collection of hierarchically classifying concepts in an automatic manner from text corpora. 
Gugnani and Hemant~\cite{gugnani2020implicit} created a taxonomy of skills in multiple fields that was mined from public online web dataset resources and then used four modules to split them (Named Entity Recognition, grammatical tagging, embedded word2vec space of skill-term, skills-term dictionary), they generate a binary probability equation that determines if the parsed item is a skill-term. The probability equations combine the models decision including ONet\footnote{https://www.onetonline.org/}, Hope\footnote{https://www.computerhope.com/} and Wikipedia dictionaries. After preparing the taxonomy skills, they use it to extract explicit skills, and the implicit skills (interpreted from similar jobs). Finally, Cosine similarity and TF-IDF were used to match skills and explicit-implicit skills.

Singh \al~\cite{singh2017propensity} used a job-role taxonomy that describes the job roles inside the organizations that typically have various job roles, where the hierarchy describes job categories and job roles at the top, until reaching the individual skills needed to satisfy the jobs category at lower levels of the taxonomy. The goal of this work is to determine the target skill that a candidate needed to learn. 
Javed \al~\cite{javed2015carotene} used the common ontologies O*NET to associate the job ads and resumes to the CareerBuilder \footnote{https://www.careerbuilder.com/} job title taxonomy.

\subsubsection*{Ontology}
An Ontology is a representation of a set of concepts within a domain and the relationships between those concepts \cite{ivanovic2014overview}.

Several approaches that are doing the matching JD/Resume chose to create their skills base ontology. Balachander \al~\cite{balachander2018ontology} built a custom technical skills ontology by crawling DBpedia and then used to compute the similarity/ dissimilarity between these features to show the relationship between skills in the ontology. Besides, Celik~\cite{celik2016towards} deployed an ontology-based resume parser (ORP) that is constructed from many domain ontologies where each ontology has its domain-based concepts, properties, and relationships according to the segments of a personal resume (education, location, abbreviations, occupations, organizations, resume). Their ORP is based on six modules that treat resumes (converter, segmenter, parser engine, normalization, classification and clustering of concepts, and generating personal résumé ontologies for individuals). The resumes are analyzed semantically using the framework and a Jaro-Winkler distance algorithm was used to reform the misspelled parsed terms. A resume ontology was proposed also by Mohamed \al~\cite{mohamed2018smart} where they considered personal information, skills, educational qualifications, certifications, and work experience. They proposed a manual update of the ontology in case the new skills feature is not recognized.

Guo \al~\cite{guo2016resumatcher} presented in their methodology RésuMatcher a system that generates a domain-specific ontology.  To compute the similarity and relationship between skills, DBpedia knowledge taxonomy was used. Corde \al~\cite{corde2016bird} created a skill ontology, where they consider the skill similarity of the job seeker and a job description by computing the path distance between two skills.

Maree \al~\cite{maree2019analysis} built a semantic network from refined concepts of job offers and resumes where words’ semantic relationships are mapped in a network. They utilize ontologies, WordNet \cite{miller1998wordnet}, and YAGO \cite{suchanek2007yago} to enrich the knowledge with semantic resources and occupational classifications. The produced networks from the resume segments were matched with their corresponding networks that are extracted from the job offer using Jaro–Winkler distance. 
The same idea was applied by Nimbekar \al~\cite{nimbekar2019automated}, where they derived the relatedness between skills from both resumes and job posts to construct a semantic network. The semantic network was used as input to the matching algorithm to measure the closeness JD/Resume.

In the context of our thesis, we will consider the ESCO ontology to be used. Since, it contains skills, competencies, qualifications, and occupations. ESCO is bridging language barriers by providing terms for each concept in 26 European languages and Arabic. To map between the different languages, each occupation, knowledge, skill/competence provides with a unique universal URI over the web.  
ESCO provides a short explanation of the meaning of the occupations and clarifies its semantic boundaries. 


\begin{figure}[H]
\centering
\includegraphics[width=130mm,scale=0.2]{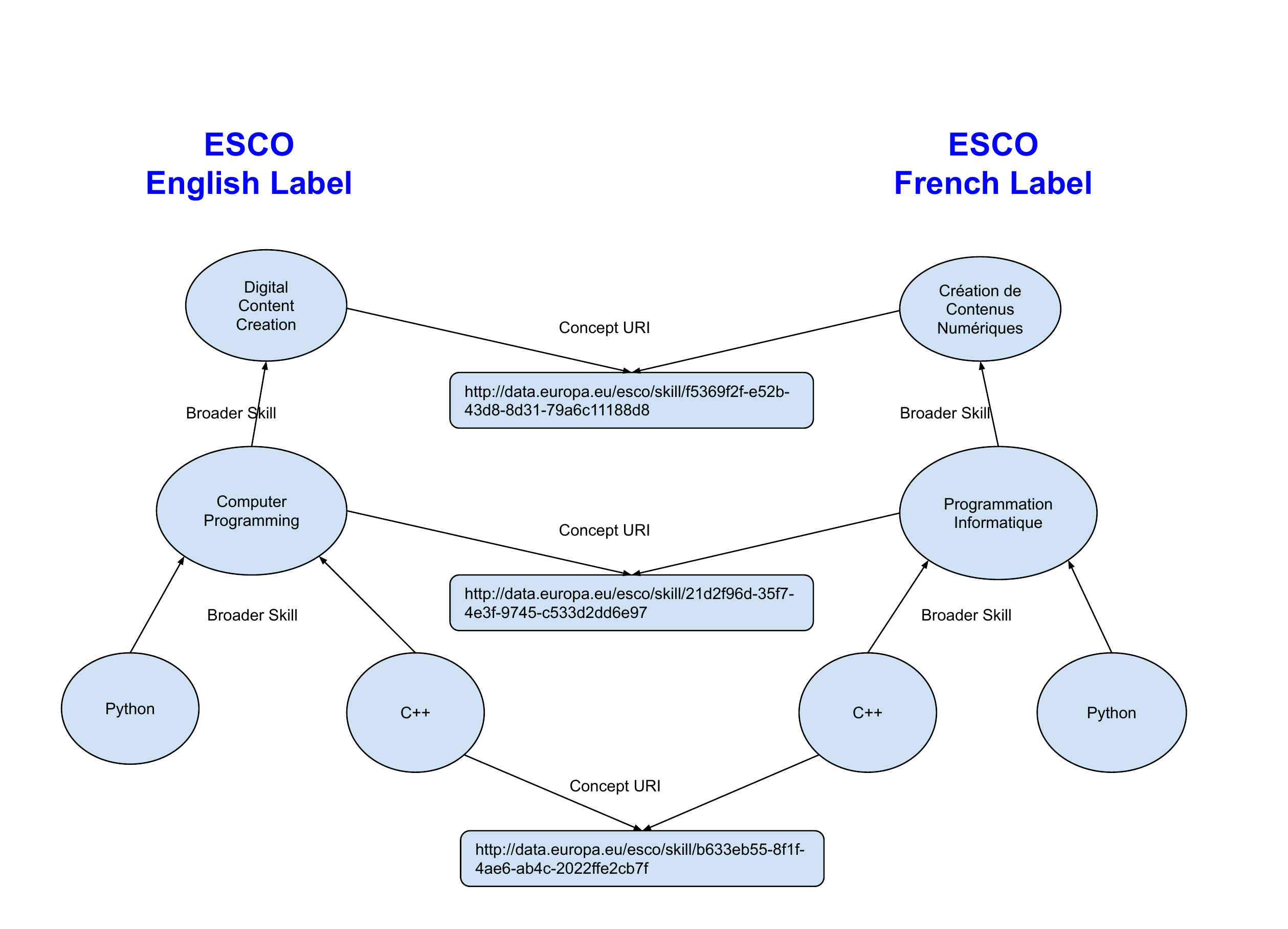}
\caption{Example of the ESCO ontology labeled with a unique URI in English and French languages}
\label{ontology}
\end{figure}

We present in Figure \ref{ontology} an example of the ESCO ontology. We can notice that the skills are listed in a hierarchy. Each skill (or ESCO concept) has a unique concept URI which is used to identify and map the same skill in different languages. 
For instance, Software Developer (EN) and  Développeur de Logiciels (FR) share the same concept URI. \footnote{http://data.europa.eu/esco/occupation/f2b15a0e-e65a-438a-affb-29b9d50b77d1}
A real-life example of mapping a resume skill to a job description required skill would be as follows: (i) If a resume skill (e.g. C++) directly matches with a skill listed in the job description (e.g. c++), it will be a perfect match. (ii) However, the model is also beneficial to map the skills indirectly. e.g., mapping can also be done when resume skill is more specific (C++) but the job description skill is broader (computer programming), or vice versa. Since, C++ is a narrower skill of Computer Programming, it will be picked up for mapping because the model connects these two as parent-child.

\subsection{Neural Network Architectures} 
\label{ml}
Different deep learning methods have been used in JD/Resume matching and advanced the performance and flexibility of solving text mining problems. Some previous studies used deep learning methods to address NLP tasks \cite{he2019career}. Among various deep learning models, Recurrent Neural Network (RNN), Convolutional Neural Network (CNN) are widely-used architectures, that can provide effective ways for NLP problems \cite{qin2018enhancing}.

\subsubsection{Recurrent Neural Network (RNN)}
\begin{figure}[H]
\centering
\includegraphics[width=87mm,scale=0.2]{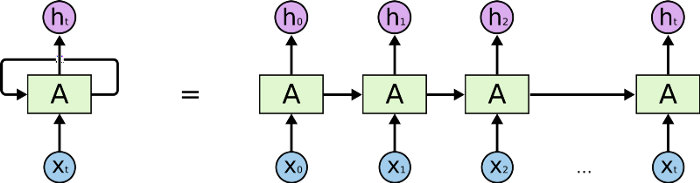}
\caption{An unrolled Recurrent Neural Network (Original figure from \cite{Understa74})}
\label{rnn}
\vspace{-5pt}
\end{figure}

Recurrent Neural Network (RNN) architecture is widely used in many NLP tasks, it is designed to process sequential information of varying lengths. An RNN performs the same task for every element of a sequence, with the output depending on the previous computations, which enables the model to predict the current output conditioned on long-distance features. Figure \ref{rnn} shows the architecture of the Recurrent Neural Network.

Qiao \al~\cite{qiao2019mlca} created a competency analysis model, where it has a job description or a resume as input and provided the job requirements and the job seekers' competency as outputs. 

RNNs have a feedback loop in the recurrent layer of the previous computation. However, it can be difficult to train them to solve problems that require learning long-term temporal dependencies, due to the vanishing and exploding gradient when computing the loss function \cite{hochreiter1998vanishing}.

The Long Short-Term Memory network (LSTMs) was introduced \cite{hochreiter1997long} which is a variation of RNN that uses special units in addition to standard units. LSTM units include a 'memory cell' that can maintain information in memory for long periods of time to understand the meaning. A set of gates is used to control when information enters the memory when it's output, and when it's forgotten.
As a variant of LSTM, Bi-directional LSTM (BiLSTM) is composed of a forward LSTM and backward LSTM \cite{graves2005framewise} that can preserve information from both past and future.

\begin{figure}[H]
\centering
\includegraphics[width=87mm,scale=0.2]{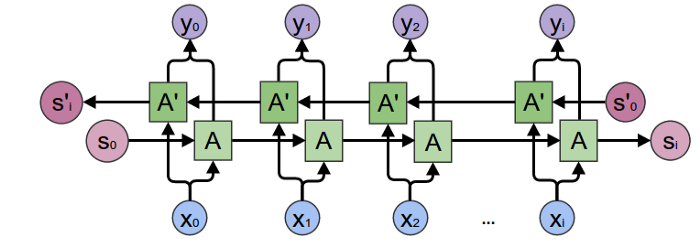}
\caption{Bidirectional LSTM architecture (Original figure from \cite{BiLSTMWh16})}
\label{biLSTM}
\vspace{-5pt}
\end{figure}

BiLSTM architecture as shown in Figure \ref{biLSTM}, has been used in different ways in the matching between resumes and job descriptions to better understand the context of the text. For example, Qin \al~\cite{qin2020enhanced} used BiLSTM to model the word-level representation of job posting and resumes. In contrast, Luo \al~\cite{luo2019resumegan} provided aggregated representation for resume experiences embedding words. However, Nigam \al~\cite{nigam2019job} used a BiLSTM to capture candidates' interactions with jobs and get an idea about their preferences by leveraging both past as well as future candidate-job interactions of the latent job preferences.

Similar to LSTM, Gated recurrent units (GRUs) is a gating mechanism in RNN to track long-term dependencies effectively while mitigating the vanishing/exploding gradient problems. The GRU operates using a reset gate and an update gate. The reset gate sits between the previous activation and the next candidate activation to forget the previous state, and the update gate decides how much of the candidate activation to use in updating the cell state \cite{cho2014learning}.

Bian \al~\cite{bian2019domain} employed a bi-directional recurrent neural network with gated recurrent unit (BiGRU) to model both sentences and documents in a job
posting or a resume.

\subsubsection{Convolutional Neural Network (CNN)}

\begin{figure}[H]
\centering
\includegraphics[width=100mm,scale=0.2]{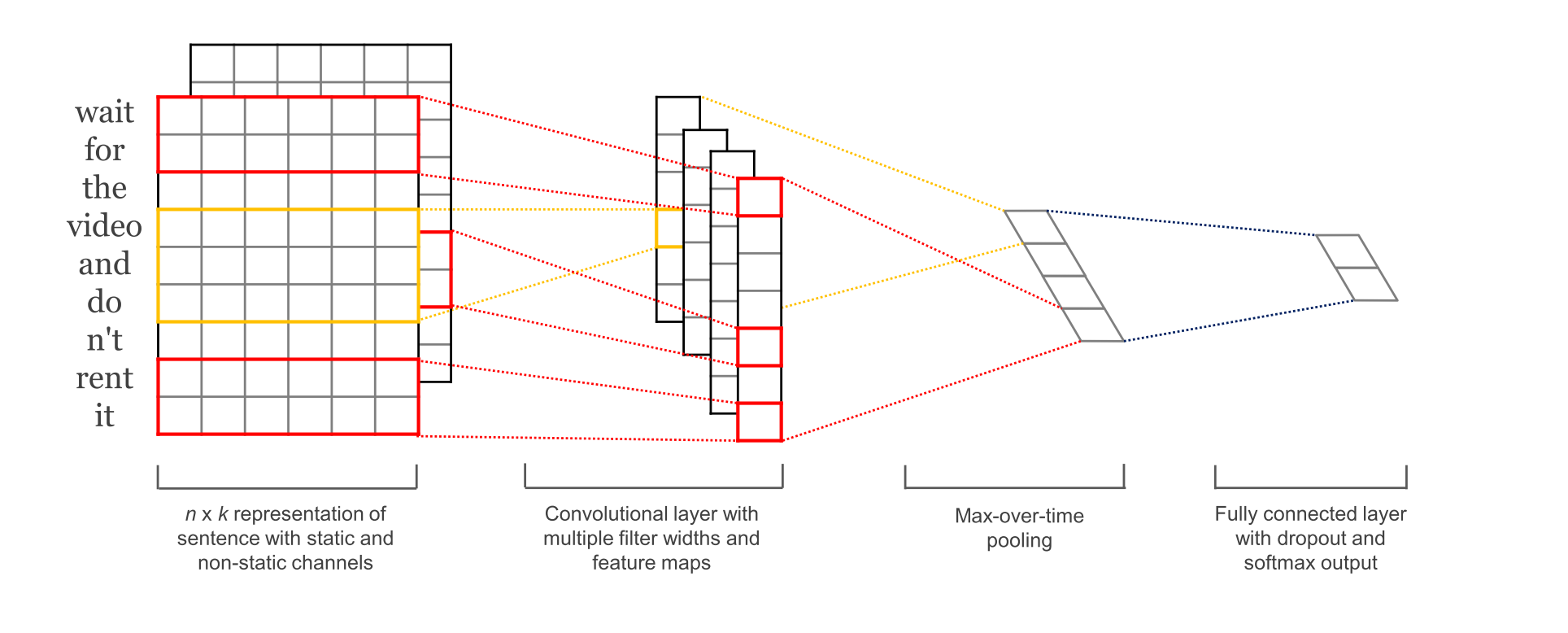}
\caption{Convolutions Neural Network architecture (Original figure from \cite{cnn})}
\label{cnn}
\vspace{-5pt}
\end{figure}

Convolutional Neural Network CNN aims at modeling hierarchical relationships and elicit local semantics. The effort of applying CNN in text mining can date back to Kalchbrenner \al~\cite{kalchbrenner2014convolutional} where they proposed a Dynamic Convolutional Neural Network (DCNN) to model sentences.
In the main use of the CNN architecture in the context of resume and job description matching is to extract the features \cite{guohao2019competency}.
Le \al~\cite{le2019towards} deployed a CNN to identify the skills and characteristics that are important for the matching between job postings and resumes. Figure \ref{cnn} shows the architecture of the Convolution Neural Network.

Zhu \al~\cite{zhu2018person} proposed a framework called PJFNN, based on CNN. They used a Person-Job Fit Neural Network to learn the joint representations of Person-Job fitness from historical job applications. They justified the use of CNN architecture rather than RNN for textual data modeling for better hierarchical relationships and local semantics between a job posting (resume) and its requirement (work experience) items. Jiang \al~\cite{jiang2020learning} used word embeddings by utilizing Zhu \al~\cite{zhu2018person} model and created a matrix for each resume where each row is a fixed-length sentence. The matrix used as input for a conventional neural network that is supposed to extract explicit features.

He \al~\cite{he2019career} presented a model to predict the career trajectory of talents based on their features inside their resumes. They classified the features on numerical and textual features; they applied one-hot encoding to represent the numerical features and word2vec to represent the textual ones. They used these datasets as input to the five layers CNN model (input layer, convolution layer with a ReLU activation, max-pooling and dropout, Dense layer, and softmax layer).

Luo \al~\cite{luo2019resumegan} used CNN in the job description. This was aimed to get fewer parameters than multilayer perception which reduced the model complexity. They represented job requirements by embeddings with an attention layer to identify important job requirements. Finally, the embeddings were fed to the convolutional neural network.

\subsubsection{Graph Neural Networks (GNN)}
Graph Neural Networks (GNN) are developed to directly learn on networks or graphs, where nodes are represented as propagated along with edges and updating node representations with the combination with its neighbors to generate node embeddings through the design of multiple graph convolution layers \cite{shi2020evolutionary}.


Bian \al~\cite{bian2020learning} proposed a predictive JD/Resume matching network that consists of a decision taken by two components that capture semantic compatibility of JD/Resume in two different views, (1) text-based matching model between JD/Resume and (2) relation-based matching model that link semantics between similar JDs and Resumes. In the text-based matching model, they represent the sentences of resumes and jobs using the BERT encoder and feed them to a transformer-based architecture where the output represents the overall document. An explicit explanation of the transformers is provided in the subsequent section \ref{tf}. In the relation-based matching model, they made a graph to represent the interaction between resumes and jobs (job-to-resume, resume-to-resume, and job-to-job) where job-to-job and resume-to-resume are linked by categories labeling and job-to-resume are linked based on keywords importance. Then, they create a matching model based on a relational graph convolutional network of the JD/Resume relation graph. After creating both models in different ways, they proposed to update the model weights using stochastic gradient descent (SGD) to penalize the instances where there is a disagreement and filtering the poor learned instances.

In a similar context, Zhang \al~\cite{zhang2018multiresolution} leveraged a graph convolutional network to match a query (short text) and a long text document. They create a keyword graph from the text document via three steps: document preprocessing with NLP method, Keyword extraction using the TF-IDF technique, and construction of the edges between the keywords. To capture the structural information, they presented a graph attention network mechanism to learn the representations and model the local interactions to handle the short-long text matching problem.

\subsubsection{Transformer architecture (Attention-based components)}
\label{tf}

In 2017, a deep learning technique was introduced known as the Transformer. In natural language processing, transformers were presented to deal with the tasks in which the data is sequential data. For example, tasks like translation and text summarization are handled using transformers. More specifically, BERT is a transformer-based model that has been used in a wide variety of NLP tasks. Most importantly, BERT has outperformed many traditional nonneural network models. 

The attention mechanism is a part of a neural architecture that was introduced to enhance encoder-decoder models to alleviate the fact that the context vector of an RNN becomes an information bottleneck. It allows focusing on certain parts of the input sequence by assigning higher values to more relevant elements when predicting a certain part of the output sequence, enabling easier learning and of high quality \cite{galassi2020attention}.

\begin{figure}[H]
\centering
\includegraphics[width=50mm,scale=0.2]{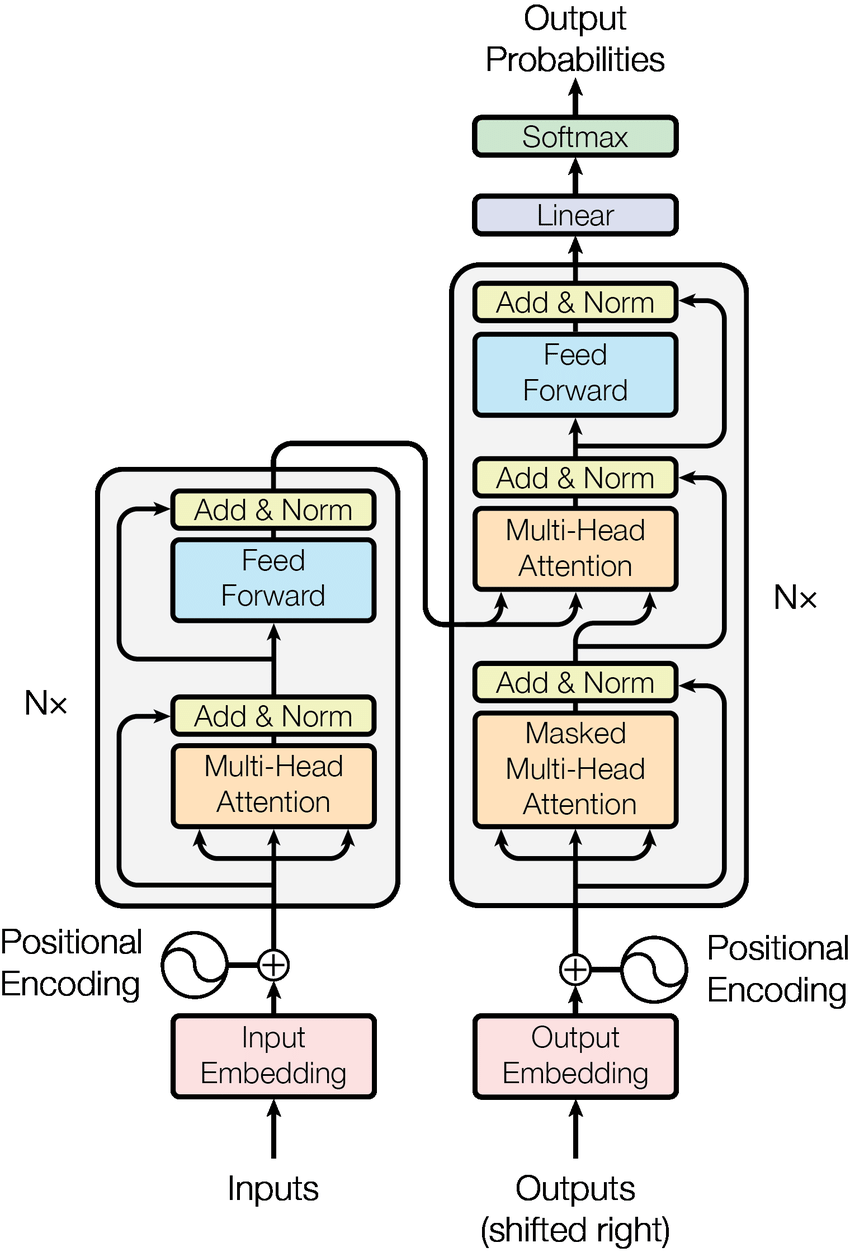}
\caption{The Transformer - model architecture \cite{vaswani2017attention}}

\label{file_evolution}
\vspace{-5pt}
\end{figure}

Qin \al~\cite{qin2020enhanced} proposed a framework called TAPJFNN to predict a person-job fit based on the topic of the job description. They applied the attention mechanism using the softmax function in different stages in their architecture for the purpose to calculate the weight of word embeddings and improve the interpretability of the relationships between job postings and resumes. They apply it to the job requirements, candidate experiences, and hidden layers of their proposed neural network. Same with Bian \al~\cite{bian2019domain}, where they applied an attention-based RNN encoder to derive the sentence representation of the job posting and a resume.

Luo \al~\cite{luo2019resumegan} used a word embedding technique to provide a meaningful representation of the words/phrase and their context in the job posting and resume. They considered three major groups of features in the resume (experience, skills, and talent field) and the recruiter job post. They adopted the hierarchy of attention architecture to assign high values ( weights) to the important features of words.

\subsubsection{Word embeddings and pre-trained language models}
Word embedding is a feature learning techniques in natural language processing where words or phrases from the vocabulary are mapped to vectors of real numbers. 
Fernandez and Suraj~\cite{fernandez2019cv} used a hybrid Average Word Embedding AWE representation of the resumes and job description. They created it by combining a trained word embedding from CV/JD with pre-trained Spanish word embeddings.

Word2vec~\cite{mikolov2013distributed} is a method to obtain distributed representations for a word by using neural networks with one hidden layer. It was implemented in multiple cases to represent job offers and profiles \cite{valverde2018job, fernandez2019cv, nigam2019job, shalaby2016entity, deng2018improved, luo2015macau, gugnani2020implicit, qiao2019mlca}. Doc2vec \cite{le2014distributed} is a feature vector representation of a document. It has been used in \cite{dehghan2020mining, gugnani2020implicit, maheshwary2018matching}. luo \al~\cite{luo2019resumegan} implementing the ELMo word embedding to represent each word/phrase in both the resumes and job-posts.

BERT is a method of pre-training language representation, where it generates multiple, contextual, bidirectional word representations. BERT only implements the transformer encoder part \cite{bian2020learning}. Different studies used BERT to solve various NLP tasks. For example,
Jiang \al~\cite{jiang2020learning} used BERT to predict the university index from a predefined university list. Devlin \al~\cite{devlin2018bert} adopt BERT as encoder layer of the sentences corresponding to the skill requirements of job post.

The following Figure \ref{bert} show an example of BERT transformer usage. The BERT was trained on different down-stream tasks and then fine-tuned to the specific task of questions/answers.

\begin{figure}[H]
\centering
\includegraphics[width=100mm,scale=0.2]{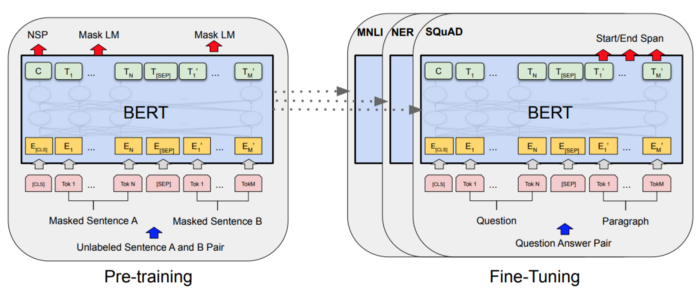}
\caption{BERT: Pre-training of Deep Bidirectional Transformers for Language Understanding architecture (Original figure from \cite{devlin2018bert})}
\label{bert}
\vspace{-5pt}
\end{figure}

In the following Figure \ref{bert_fine_tuning}, a simplified overview of the way that a language model can be pre-trained on a large text corpus, and then fine-tuned using a dataset for a specific task.

\begin{figure}[H]
\centering
\includegraphics[width=120mm,scale=0.2]{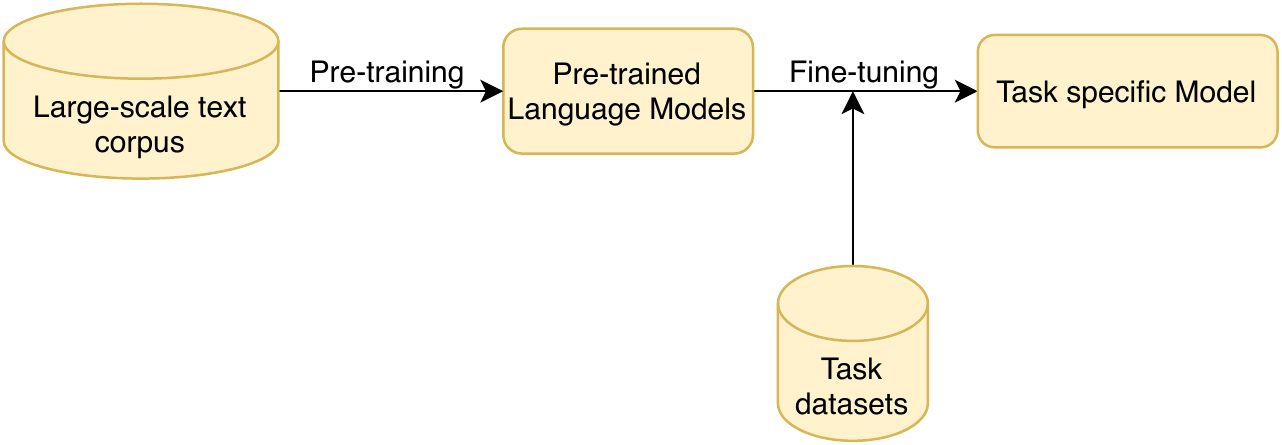}
\caption{Overview of fine-tuning pre-trained models}
\label{bert_fine_tuning}
\vspace{-5pt}
\end{figure}

\subsubsection{Classical Machine Learning} 
In several works of JD/Resume matching in literature have been using classical machine learning, Logistic Regression (LR), Factorized Machine (FM), GBM (gradient boosting), RF (random forest), VOBN (variable-order Bayesian networks), SVM (support vector machine), C4.5, Naive Bayes, Adaboost (AB), Random Forests (RF), Gradient Boosting Decision Tree (GBDT), Linear Discriminant Analysis (LDA), and Quadratic Discriminant Analysis (QDA), as base models in their experiments to validate the performance of their proposed architecture models to solve the JD/Resume matching \cite{luo2019resumegan, pessach2020employees, zhu2018person,malherbe2014field,nigam2019job}.

In a particular case, Ozcan and Oguducu \cite{ozcan2016applying} applied different classification techniques in the job recommender system to deal with the cold and non-cold start of candidates recommendation. Qin \al~\cite{qin2020enhanced} created a mean vector of word embedding vectors of ability requirements and candidate experiences to run their experiments on the classical machine learning (LR, DT, AB, RF, GBDT).

\subsection{Multilingual matching models} 
\label{lingual}

\begin{table}[H]
\centering
\caption{Recommendation base multilingual matching models}
\label{multilingual_table}
\begin{adjustbox}{max width=12cm}
\begin{tabular}{|c|c|c|c|c|}
\hline
 & celik\cite{celik2016towards} & malherbe\al~\cite{malherbe2016bridge} & Tamburri\al~\cite{9191408} & Shakurova \cite{shakurova2019best} \\ \hline
Method used & \begin{tabular}[c]{@{}c@{}}Ontology based \\ parsing of resumes\end{tabular} & \begin{tabular}[c]{@{}c@{}}create graph\\  of skill base knowledge \end{tabular} & \begin{tabular}[c]{@{}c@{}}identify skills from \\ parsed resumes\end{tabular} & \begin{tabular}[c]{@{}c@{}}bilingual dictionary \\ from parsing CV\end{tabular} \\ \hline
Learnt Language & labeled (English + Turkish) & French \& English & English & English \\ \hline
Test Language & english \& Turkish & French \& English & Dutch-Flemish & \begin{tabular}[c]{@{}c@{}}English/German \& \\ English/Dutch\end{tabular} \\ \hline
Match based on & \begin{tabular}[c]{@{}c@{}}ontology labeling\end{tabular} & \begin{tabular}[c]{@{}c@{}}Skills extracted from \\ JD/candidates\end{tabular} & \begin{tabular}[c]{@{}c@{}}Skills lines \\labeling \end{tabular}& cross-lingual learn \\ \hline
\begin{tabular}[c]{@{}c@{}}Experiment\\  (resume/job)\end{tabular} & Only one resume case study & 100 Jobs & 10K jobs & \begin{tabular}[c]{@{}c@{}}Compare (200 \\ and 500 CV)\end{tabular} \\ \hline
\begin{tabular}[c]{@{}c@{}}Precision(P)/Recall(R)\\extraction\end{tabular} & - & P(0.81) & P(0.73) / R(0.94) & 0.81 (f1 measure) \\ \hline
External knowledge & - & \begin{tabular}[c]{@{}c@{}}French/English DBpedia, \\ StackOverFlow tags\end{tabular} & \begin{tabular}[c]{@{}c@{}}Pre-trained\\ Bert\end{tabular} & \begin{tabular}[c]{@{}c@{}}Pre-trained\\ embedding layer\end{tabular} \\ \hline
Features used & \begin{tabular}[c]{@{}c@{}}Skills, education,\\ location, occupation,\\ concepts, organisation \end{tabular}& Skills & Skills & Skills \\ \hline
year publication & 2016 & 2016 & 2020 & 2019 \\ \hline

\end{tabular}
\end{adjustbox}
\end{table}

Multilingual systems have received the special attention of many researchers. Multilingualism plays a very important role in these systems because we don’t have one language to deal with, in regard to resumes and job-seeking activities \cite{baaij2012eu}. Resumes to job description matching systems in a are difficult to find especially when they have to deal with a single language. To figure out interpretable figures, and make the best use of multilingual text, one major focus of the project is to derive useful and semantic knowledge. Table \ref{multilingual_table} show the recommendation base, multilingual matching models.

Lexicon differs with respect to the languages and learning the lexicon from one language to another is a difficult task  \ie{English lexicons are different than Turkish} \cite{bal2016matching}. Bal \al~\cite{bal2016matching} analyzed Turkish job advertisements and tried to identify the structure of the common sentences to create some rule patterns to extract. They addressed the difficulties encountered such as the Turkish verbs are at the end of the sentence, whereas in English, verbs are at the middle of the sentence. Multiple solutions for JD/Resume matching were proposed to treat several individual languages in the literature \ie, Indonesian \cite{kusnawi2019decision}, Vietnamese \cite{nguyen2016adaptive}, Brazilian \cite{valverde2018job}, Belgium \cite{reusens2018evaluating}, Spanish \cite{fernandez2019cv}, Chinese \cite{deng2018improved}. Cross-lingual may be a solution.

Cross-lingual embedding represents words in multiple languages, they are crucial for task scaling of multiple languages by transferring knowledge from high resource languages \ie English to low resource language \ie German and Dutch \cite{shakurova2019best}.
Lena \al~\cite{shakurova2019best} evaluated the embeddings that are on the sequence labeling tasks of parsing CV and they also show the size of a bilingual dictionary, frequency of dictionary words along with performance measures. The researchers have conducted an experiment on Dutch-English and German-English cross-lingual embeddings. They used Canonical Correlation Analysis (CCA) linear projection in a monolingual vector to estimate Dutch/German embeddings in terms of English space. Subsequently, these embeddings are used for the sequence model. The sequence model is always trained in terms of English space. The data used for training is either English data along with the Dutch/German training data or just English training data. Once the model is created to support multilingualism, it is tested Dutch/German test data. Several other factors have been considered for experimenting with bilingual dictionaries like size, data source, and bilingual dictionary frequency.

The French and English multilingual aspect was considered by Malherbe and Aufaure \cite{malherbe2016bridge} proposed a knowledge base architecture (English and French) of skills that is extracted (job offer, resumes) from the job offer website. The crawled data was filtered to include skills data based on the term frequency in the corpus. After formalizing these documents, the system is coupled with external sources of information that are named as Stack-Overflow and DBpedia. A normalization technique was applied to associate each job offer and profile with the corresponding skills base. The link was based on having the same alias between the tags of Stack-Overflow and the concepts in DBpedia. The very 1st step was to process the skill terminologies obtained from English and French sources for which a hypothesis “to use an expression that appears frequently in the skills field content of the profiles” was proposed. The 2nd step of the framework is based on extracting the related concepts in specially chosen knowledge graphs, namely DBpedia and Stack-Overflow tags. Most of the data uploaded by a candidate are in unstructured text form. Their extraction of multilingual skills using their proposed approach was able to reach 80\% \cite{malherbe2016bridge}.

Celik~\cite{celik2016towards} presented a semantic-based extraction system for matching resumes for any job opening to gather important information from the resumes. The study proposed an ontology-o=based resume Parser (ORP) system for Turkish and English language used in resumes along with concept-matching tasks, the data is analyzed semantically, and then it is parsed with important information like education, experience, business, and features.  The study was divided into 6 major steps, starting with the conversion of the resume, partitioning it into segments, then parsing important data from input, normalizing that data, after that applying clustering and classification tasks to focus on the important sections of the resumes. The framework used is based on SWRL and provides a formal OWL ontology with rules in the abstract syntax. Jaro-Winkler distance algorithm was used to detect incorrectly spelled words. Added to that, they ensure multilingualism in this research by translating items from the different ontologies and assigning tags labels of English or Turkish languages.




Tamburri \al~\cite{9191408} presented a DataOps model based on agile practice in skills extraction from resumes and jobs, featuring machine learning models. The researchers have applied a DataOps pipeline for skills extraction constructed from five steps (Data Pre-processing, Sentences Annotation for Learning, Model Training, fine-tuning training, and prediction). Their experiment is directed on using BERT cloud base-model pre-trained in English \cite{pires2019multilingual} and fine-tuned with annotated sentences extracted from vacancies. They applied their model to a Dutch-Flemish cross-border labor market for job seekers. 


Linkedin proposed a method to make users profiles in other languages and create another profile in the preferred language. Moreover, users can set the language that their profile will be displayed, and Linkedin does not translate the content or messages. People viewing a profile can choose from the language a profile owner setup before \footnote{https://www.linkedin.com/pulse/20140710185825-25298675-multilingual-create-a-secondary-language-profile-on-linkedin/}.

\textbf{To summarise}, with the growth of powerful pretrained language models \ie BERT, the need to fine-tune in a specific field or even in a specific required language to train such model to identify the features \ie skills. To do so, a dataset needs to be labeled to fine-tune a pretrained model. Tamburri \al~\cite{9191408} used a specific dataset gathered from parsing JD/Resume and labeled manually via a domain expert and this labeled dataset became the standards dataset used in the fine-tuning. However, such sources are not enough to cover enough knowledge, also, expensive for manual labeling. In our proposed method, the use of reach multilingual ontology/knowledge base is required to be used in labeling JD/Resume for fine-tuning BERT language model purposes.

\subsection{Biases in the automated e-recruitment Machine Learning algorithms decisions}
\label{biases_ml}

The machine learning algorithms used in automated programs of hiring usually train their models on a pair of matched JD/Resume. These algorithm decisions have been shown a bias decision \ie{cognitive bias, coworkers, demographic subgroup, etc} \cite{langenkamp2020hiring}. For example, Amazon created a hiring tool in 2014, that has to parse resumes and infer the best candidates using their own workforce over the past 10 years, where they trained their models on a large majority of its existing workforce of white and male persons \cite{mac2014amazon} which lead to systematical bias against female applicants \cite{goodman2018amazon}.

Pessach \al~\cite{pessach2020employees} consider dealing with the biases that may happen in the e-recruitment prediction model. They attentively consider including a dataset that represents a wide range of the heterogeneous populations through a mathematical programming model. 

In this thesis, we will consider dealing with the potential biases we may have in the model by considering a wide range of the JD/Resume per topic in the training phase, including various diversified features. A collaboration with Airudi will open access to a wide range of multivarious datasets (JD/Resume). 

\subsection{Data and Machine Learning traceability}
\label{traceability}

In the JD/Resume matching workflow, common steps are realized (parsing, feature extraction, model training, experimentation, and validation) and repeated over time. 
Since JD/Resume matching models and workflows are centered around data, it is important to keep track of the data used at each step, machine learning models, and iteration of the workflow, i.e., recording the history of the different ML stages, to ensure the reproducibility of the ML pipeline (same inputs, same outputs) and to track data provenance. JD/Resume matching pipelines need to be automatically tracked in a way that guarantees that all the files and metrics will be reproducible or fetch the full context of an experiment or to perform a new iteration.

The versioning of ML data and models is a young and growing practice, with several tools created to help developers with tracking the various aspects of their workflow. Airflow~\cite{airflow} is used to create, schedule, and monitor machine learning workflows as a directed acyclic graph (DAG) that may be composed of multiple tasks. Similarly, Luigi~\cite{luigi} is a workflow engine framework that helps to write static and fault-tolerant data pipelines in Python. Miguel \al~\cite{miguelmarvin} presented the Marvin engine that supports the exploration and model development of distributed computing systems for data-intensive applications. It provides a standard interface to allow other applications access to shared model artifacts and to support high throughput and processing of large datasets. MLflow~\cite{MLflow} is a platform to streamline machine learning development. It  is divided into three components for (1) tracking experiments, (2) packaging the code into reproducible runs, and (3) sharing and deploying models trained using diverse ML frameworks. Kedro~\cite{kedro} provides a development workflow framework that implements software engineering best-practice for data pipeline construction, basically leveraging data abstraction and clear code organization to bring models into production. Pachyderm~\cite{Pachyder} is a data science platform aimed at an enterprise that combines data lineage \cite{atwal2020dataops} with end-to-end pipelines on Kubernetes, with a graphical pipeline builder and data versioning.

DVC~\cite{dvc} is a data/model versioning tool that is integrated with git repositories, such that the history of data, models, and code can evolve together in an efficient manner. It is designed to handle large files, data sets, machine learning models, metrics, and code. DVC was designed to support the gradual adoption of ML capabilities in traditional software projects. 

In our thesis, we will consider studying the need to propose system traceability based on one or more versioning tools. Therefore, studying the best practices of applying such tools in software repositories is required. A study in this context has been proposed and sent to SANER2021.

\section{Research Methodology}
\begin{figure}[H]
\centering
\includegraphics[width=60mm,scale=0.2]{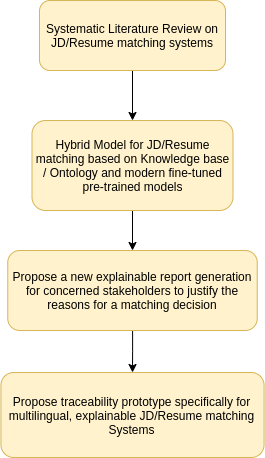}
\caption{Overview of the Research methodology of the Thesis}
\label{methodology}
\end{figure}

We will describe in this section our methodology and steps that we will follow to address the following research questions:

\begin{itemize}
\item[RQ1:] \textbf{\RQOne}
\item[RQ2:] \textbf{\RQTwo} 
\item[RQ3:] \textbf{\RQFour}
\item[RQ4:] \textbf{\RQFive}
\end{itemize}\leavevmode\newline
We propose a general methodology to achieve the thesis goal shown in Figure \ref{methodology}.

\subsection{\RQOne}

As a first objective, we created a systematic literature review to address the following requirements related to JD/Resume matching (methodology and preliminary results are shared in Section III):
\begin{itemize}
    \item Understand the challenges of JD/Resume matching.
    \item Identify the features used in the literature that may help to make the matching.
    \item Categorize the different methodologies to extract features from JD/Resume used in literature.
    \item Categorize the different methodologies for matching between resumes and job descriptions.
    \item Identify the different metrics used to evaluate the matching process.
\end{itemize}

\subsection{Overview of the Proposed Architecture}
\begin{figure}[H]
\centering
\includegraphics[width=140mm,scale=0.2]{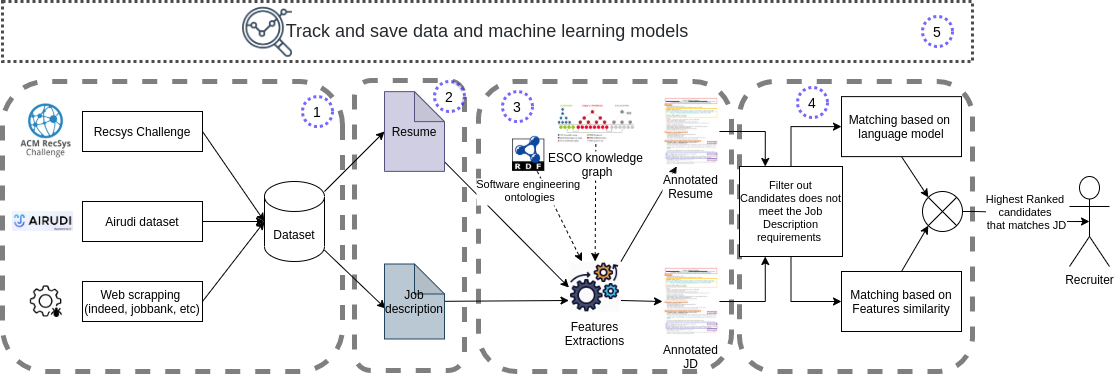}
\caption{Overview of the proposed architecture of matching Resumes to the Job description}
\label{overview}
\end{figure}

During our PhD, we propose the architecture presented in Figure \ref{overview} as a pathway to be proven to improve the matching between jobs and resumes. The proposed architecture is supposed to resolve the thesis research questions related to the JD/Resume matching:
\begin{itemize}
    \item \RQTwo
    \item \RQFour
    \item \RQFive
\end{itemize}

The proposed architecture of matching Resumes to the Job description is based on processing the following points:
\begin{enumerate}
    \item Dataset sources and pre-processing
    \item Resume and job description features
    \item Features extractions
    \item The matching system
    \item Traceability \& Explainability of the matching system

\end{enumerate}

\subsection{Data Sources and pre-processing}
In this section, we describe the different sources and alternatives of the dataset and the possible preprocessing steps that may be applied.
\subsubsection{The Airudi dataset}
During our internship, we received ethics approval to have access to the company databases of resumes and Job descriptions of our industry partner Airudi. Such a dataset is considered critical and the most important part to initiate this project since companies do not share such confidential data (resumes of persons who applied to their proposed jobs).

Airudi collects their dataset from its clients (companies looking to hire effective persons), where they apply an anonymization protocol of the candidates and hide their personal information (first name, birth date, etc). 

The provided dataset contains French and English resumes and job description, and contains the following features:

\begin{itemize}




\item Resume features: education, degree, university, work experience(skills, professional competencies, personnel competencies).

\item JD features: the job title, post description, requirements (required education, required skills, required experience), job responsibilities, and the job advantages.

\item List of matched JD/Resume that is manually constructed from recruiter clients (gold standard) with different recruitment processes \ie{accepted, refused, waiting for candidate decision, send offer to candidate}.

\end{itemize}

The dataset provided by Airudi requires preprocessing steps. We applied Regex functions for impurities cleaning, for example  blank areas and extra boisterous characters. We also fixed typos related to french language \ie{è, oe}.

We transformed the list of recruitment process information between candidates and the jobs to a binary classification (match or unmatch). For example, we assigned a match in the status of “Accepted Jobs Skills” and an unmatch for the case of “Not retained - Physical interview”. There were other cases where it is not possible to determine the matching status such as “interested candidate” or “stopped process”. We removed the empty cells from the dataset that were missing a job description or a resume. We recorded the remaining cases number in Table \ref{numbers-data}.

\begin{table}[H]
\caption{Dataset labels distribution, the relation between jobs and resumes are splitted into (unknown, match and unmatch) liaison}
\label{numbers-data}
\begin{adjustbox}{max width=11.2cm}
\begin{tabular}{|l|l|l|l|l|}
\hline
Data cells & Unknown label:0 & match:1 & unmatch:2 & total \\ \hline
\multicolumn{1}{|c|}{\textless{}jobID, CandID, match-status\textgreater{}} & \multicolumn{1}{c|}{38839} & \multicolumn{1}{c|}{7656} & \multicolumn{1}{c|}{21033} & \multicolumn{1}{c|}{67528} \\ \hline
\end{tabular}
\end{adjustbox}
\end{table}

We classified (38839/67528) 57\% of the cases as unknown, most of them are noted as "stopped interviews process" (35446 cases).
In the following, we consider only the labeled dataset as (match, unmatch) with a total of 28689 pairs of <job, resume>. We made an additional filter of cases, where the same job exists with different IDs (1576/2500 jobs). Similarly, we removed duplicates where the same job is assigned to the same candidate having different IDs. We removed cases (8 matches) having contradictory recruitment processes \ie "candidate accepted" and "candidate refused". We filter out candidates having less than 50 word (771 candidates), all the removed candidates were confirmed and verified manually. Table \ref{unique-data} show details related to the remaining jobs and resumes.

\begin{table}[hbt!]
\caption{Dataset labels distribution, the relation between jobs and resumes are splitted into (unknown, match and unmatch) liaison}
\label{unique-data}
\begin{adjustbox}{max width=11.2cm}
\begin{tabular}{|c|c|c|c|}
\hline
 & Unique total & Unique Match label & Unique Unmatch label \\ \hline

candidates & 16194 & 5776 & 12107 \\ \hline
jobs & 909  & 872 & 472 \\ \hline
\begin{tabular}[c]{@{}c@{}}total pairs\\ <candidate, job>\end{tabular}  & 27257 & 7090 & 20167 \\ \hline
\end{tabular}
\end{adjustbox}
\end{table}

    
\subsubsection{Websites scraping}

We will proceed to have additional datasets from public websites, to enrich the actual dataset with the updated skills and nowadays requirements. We will consider adding job description (JD) in both languages French and English, for example, from the official Canadian job bank \footnote{https://www.jobbank.gc.ca/} or public job posting website \footnote{https://ca.indeed.com/}.
Such a public website represents a rich source of job offers, where the job poster describes his offer in two languages French and English. The Jobbank website is designed to filter the jobs by features (province, city, posted date, full/part-time, period of employment, salary, years of experience, job source, education or training, language, employment group, job categories). 

Moreover, there is another section recommending top related job categories. A list of skills and knowledge are provided to job seekers, so they can fill a list of skills they have and knowledge. Such a dataset is the reach of real and updated features. 

The indeed website offers the possibility to find candidates that are appropriate to a specific query that can be turned to advanced search \footnote{https://resumes.indeed.com/advanced-search} to include (keywords, work experience, specific experience with required years of experiences, education institution with the study field). There is an additional option (not for free) to download the job seeker profiles. 

Starting from that purpose, we plan to assign job seekers to a job description based on the following scenarios:

1- Choose a job description and via the advanced candidates research, we can specify the different fields and consider the best ranked profiles that indeed engine search proposes as our ground truth that should be assigned to the job, and the less ranked resumes as non match.

2- We can search for multiple candidates using job title or skills, then using the candidate work history, we can find the job description that he was assigned to in the past.
\subsubsection{RecSys Challenge 2017}
The challenge dataset contains user profiles, job postings, interactions that users performed on job posts, and interaction of the recruiter to the users profiles. It also contains the user job impressions, i.e., information about job postings that were shown to users. The total dataset includes 1,367,057 users and 1,358,098 jobs. Users and jobs were described by several similar attributes such as job categories, career level, industry, location, etc. In addition, the users have the educational background and details about work experience. The dataset was carried out through an anomization procedure \cite{rsys2016}.


\subsubsection{Common data pre-processing}
A data preprocessing analysis and cleaning verification should be realized. The language model should receive a clean text as input since it is performed to learn the context of a given text. The Word Sense Disambiguation (WSD) can be used to specify the correct sense for terms used in resumes and jobs description (\ie typos, missing values, etc) according to its surrounding textual content. The WSD module is integrated with NLTK python API \footnote{http://www.nltk.org/howto/wsd.html}, where it can be linked with the WordNet lexical database. Moreover, we will attempt to eliminate bias when addressing the three research questions of the thesis. To identify the dataset quality against biases, a detailed analysis to categorize the features is necessary. For example, the consideration or not of a feature (age) can increase or decrease the cases of bias during the execution of an existing matching algorithm.



\subsection{Resume and job description features}
In this section, we show the features that characterise both resumes and jobs.
\subsubsection{The resume features}
\begin{figure}[H]
\centering
\includegraphics[width=8cm,scale=0.17]{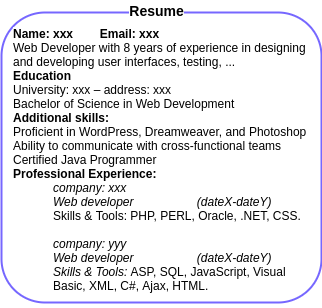}
\caption{An example of a web developer Resume}
\label{candidate_exemple}
\end{figure}

The resume is generally composed essentially of at least 3 sections: (1)Name and contact information; (2) Education (degree, school name, certifications, awards); (3) Work experience (company name, job title, accomplishments, period).\\
In a resume, we may find additional information that usually occurs: (1) Career Summary or Objective Statement; (2) knowledge or hard skills; (3) Language skills; (4) Attitudes and values and (5) others (nationality, gender, marital status, etc). Figure \ref{candidate_exemple} shows the different sections a job description can have.

\subsubsection{The job features}
\begin{figure}[H]
\centering
\includegraphics[width=8cm,scale=0.17]{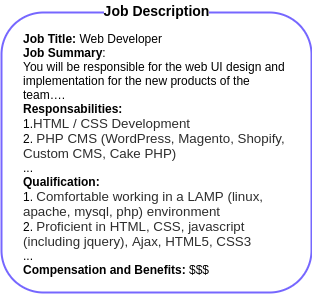}
\caption{An example of a job description for a web developer}
\label{job_exemple}
\end{figure}

The job description is  composed of a main 3 sections: (1)Job title; (2) Job responsibilities; (3) Qualifications and skills.\\
There is additional information that may be included in a job description: (1) Job summary (overview of your company and expectations for the position) (2) Compensation and benefits and (3) others (gender, age range, marital status, location, etc).
Figure \ref{job_exemple} show the different section a job description can have.

\subsection{Features extractions}
\begin{figure}[H]
\centering
\includegraphics[width=8cm,scale=0.17]{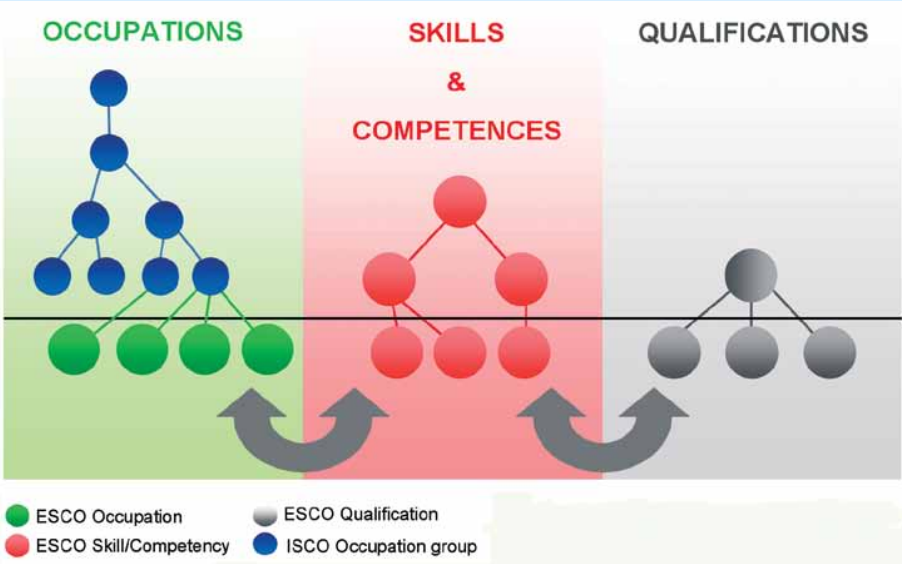}
\caption{The hierarchy structure of the ESCO ontology \cite{Presenta38}}
\label{ESCO_schema}
\end{figure}

There are different features in a resume and a job description, we propose a feature extraction method that is based on the ESCO (European Skills, Competences, qualifications, and Occupations) ontology. The ESCO ontology is a multilingual classification system of European that considers three pillars: (1) Occupation; (2) Skill (and competences) and (3) Qualification. Figure \ref{ESCO_schema} shows hierarchy structure of the ESCO ontology. 

The purpose of this section is to find a way how we can disambiguate a part from a text that describes the information in the Resume/Job, and how to relate it to a concept which is described in a knowledge base (ontology).

Since the ESCO ontology has linked pillars that describe occupations, we will proceed to interpret semantically the occupation that describes the closest possible JD/Resume based on the ontology's occupation. The ontology occupation has a short explanation of the occupation’s meaning and a specific clarification of its semantic boundaries.

However, the ESCO ontology is not perfect to cover all the fields and their interactions. The O*Net ontology and knowledge bases such as DICE have been used in several related studies \cite{maree2019analysis, zaroor2017hybrid} to identify the \textbf{skills feature}; O*Net ontology is the US's primary source of occupational information and it covers more domain \ie(medical and artistic), but other skills acronyms may not be recognized that may be covered with lexical ontology WordNet or YAGO3. Moreover, to cover more domains, additional sources of knowledge (tags from StackOverflow, quora, etc) may be added using the associations' rules (same as related to)\cite{maree2019analysis}. Although, expanding ontologies knowledge bases and labeling different multilingual categories will be a future work, due to the task complexity challenges.

The occupation pillar organises the occupation concepts in ESCO. It uses hierarchical relationships between them, metadata as well as mapping to the International Standard Classification of Occupations (ISCO) to structure the occupations.

\begin{figure}[H]
\centering
\includegraphics[width=8cm,scale=0.17]{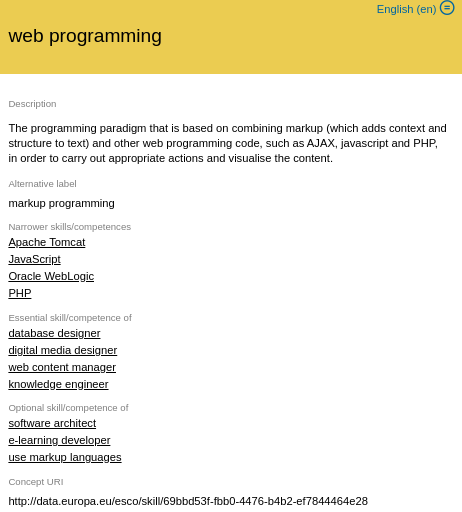}
\caption{Example of URI of Web Developer occupation in ESCO ontology \protect\footnote{http://data.europa.eu/esco/skill/69bbd53f-fbb0-4476-b4b2-ef7844464e28}}
\label{ESCO_web}
\end{figure}

Each occupation concept contains alternative labels terms that have the same meaning of the main concept and hidden terms in each of the ESCO languages. Each occupation also comes a description, scope note and definition. Furthermore, they list the knowledge, skills, and competences that experts considered relevant terminology for this occupation on a European scale \cite{Presenta38}. Figure \ref{ESCO_web} show a shorted example of a web developer occupation in the ESCO ontology.

The ESCO system presented in its last version "ESCO v1.0.8" published in August 2020 provides a web service API to make queries on the ontology. They also provide an RDF (Resource Description Framework) format of their data to be downloaded, such a format will be useful during the mapping of the Resume/JD to the closest occupation.

\subsubsection{Occupation mapping using deep contextualized word embeddings}
In this subsection, we describe our proposition to a possible way for semantic mapping using word deep conceptualization.

A query using the job title or the resume career on ESCO API can result in an exact match or to a list of possible occupations that may match with. In the case of nonexistence of the job title, the resume career, or even there is no exact match, we need to run a deeper semantic mapping.

To compare semantically between two textual parts (concept from the ontology and section parts of the JD/Resume), the contextualized word embeddings can be used to capture the word semantics in different contexts.

Language models \ie BERT, ELMo, have been used for transfer learning in several natural language processing applications. In our case, we will use transfer learning to extract the knowledge embedded in a pre-trained machine learning model. For example, the BERT model takes in consideration three aspects to represent a word and keep its meaning inside a phrase, (1) Position Embeddings to express the position of words in a sentence; (2) Segment Embeddings to distinguish between different input sentences \ie pair sentences for a matching purpose; (3) Token Embeddings that represent the word pieces vocabulary \cite{devlin2018bert}.

We plan to investigate the contextualized word embeddings from language models and compare the similarity \ie cosine distance, between the ontology occupations concepts and the section parts inside a JD/Resume. This process can be applied in occupation mapping, skills and knowledge validation. 

Depending on the task and the context, we need to run an experimentation to compare and explore the most appropriate word deep contextualized language models to our context.








\subsubsection{Feature extractions from Resumes}
\begin{figure}[H]
\centering
\includegraphics[width=12cm,scale=0.17]{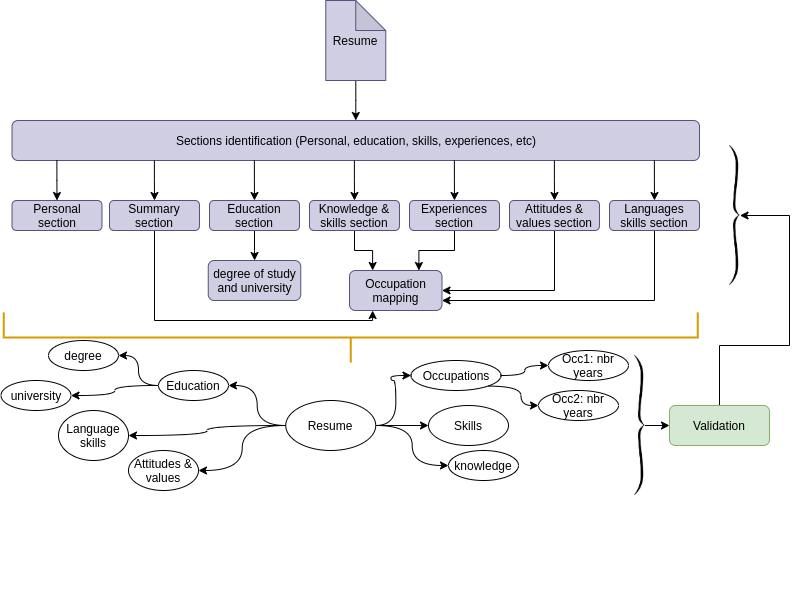}
\caption{Extracting of Candidate features}
\label{candidate_extraction}
\end{figure}

A resume is usually divided into separate sections, where every section has a message to present. To extract features, we need to start by separating these sections by using, for example regex functions. Figure \ref{candidate_extraction} shows the sections that may be identified in a resume. The purpose is to create a \textbf{graph profile} of a resume candidate that contains all the useful information inside the resume. Once the sections identified, we can use the \textit{experience section} and link it to the possible occupations the candidate may have during his working curriculum. \textbf{The occupations can be validated With the help of other sections (knowledge and skills, summary)}. In the \textit{education section}, we need to determine the degrees and the university of the candidate. To unify the degree terminology, we may use the lexical ontologies \ie WordNet or YAGO3. For example, if you search for the word "phd", the wordnet returns the different lexical ways it may be written, including hyponyms, hypernyms, and derived words.
From the \textit{summary section}, we may determine attitudes and values where we can be linked to the attitude and values ESCO taxonomy. It describes individual work styles, preferences, and behaviour. After this step, we should have a graph profile of a candidate.

\subsubsection{Feature extraction from Job Description}
A job description can be represented by an occupation in the ESCO ontology, where it can be validated with different sections (job title, job summary, responsibilities, qualifications, etc). As shown in Figure \ref{ESCO_web}, where a representation of a web developer is presented, we can assign the following mapping to job description using semantic similarities of deep conceptualized word embeddings, (1) the job summary with the job title can be linked to the description of the occupation; (2) the required responsibilities can be linked to the essential skills; (3) the required qualifications can be linked to the essential knowledge; (4) attitude and values; (5) the language skills. 

Similar to the resume, we can determine the requirements of the education degree/ university. 

\begin{figure}[H]
\centering
\includegraphics[width=12cm,scale=0.17]{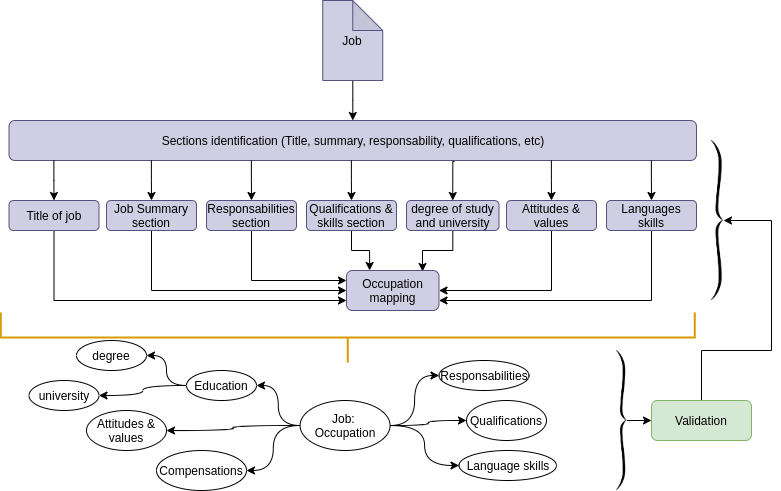}
\caption{Extracting of Job features}
\label{job_extraction}
\end{figure}

\subsubsection{Features Extraction Validation}
After extracting the features and making the graph profiles that represent the sources of JD/Resume, we will validate the feature extraction results by ensuring if the JD/Resume graph profile meets the knowledge, skills that describe the JD/Resume. An additional manual validation can be applied to a sampled case of the dataset.

\subsubsection{Language model for annotating features}
To reduce the cost of feature extraction, we plan to automate this task once we have enough labeled dataset, by applying the advanced existing pre-trained language models \ie BERT and fine-tune it to determine the different entities (dataset labeling) in resumes and job descriptions \cite{li2020survey}. \\

\subsection{\RQTwo}
\begin{figure}[H]
\centering
\includegraphics[width=12cm,scale=0.17]{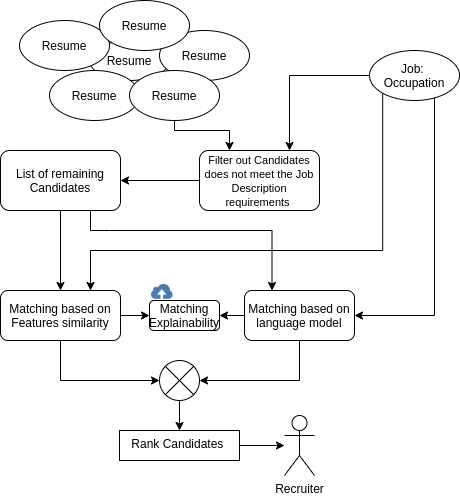}
\caption{Proposed matching system}
\label{matching_model}
\end{figure}

During this section, we present the matching system that is based on (1) deeply textual matching using language model transformer; (2) filtering out the non-conform candidates using feature similarity; (3) feature matching models.
Figure \ref{matching_model} shows the different parts of the proposed matching system.

\subsubsection{Baseline model: Job-Resume matching based on language model transformers}
We designed a baseline model to match resumes with job description. The model will be used as a reference point to evaluate the evolution of the proposed architecture performance. The proposed model is based on a BERT transformer that aims to classify a pair of <job, resume> as a match or not. BERT is a powerful language understanding model. It is characterized by a deeply bidirectional text learning, and that it can use a large amount of plain text data for training. However, BERT has a length limit on the input text, e.g., 512 words, which prevents the accurate modeling of long documents.

\begin{figure}[H]
\centering
\includegraphics[width=8cm,scale=0.17]{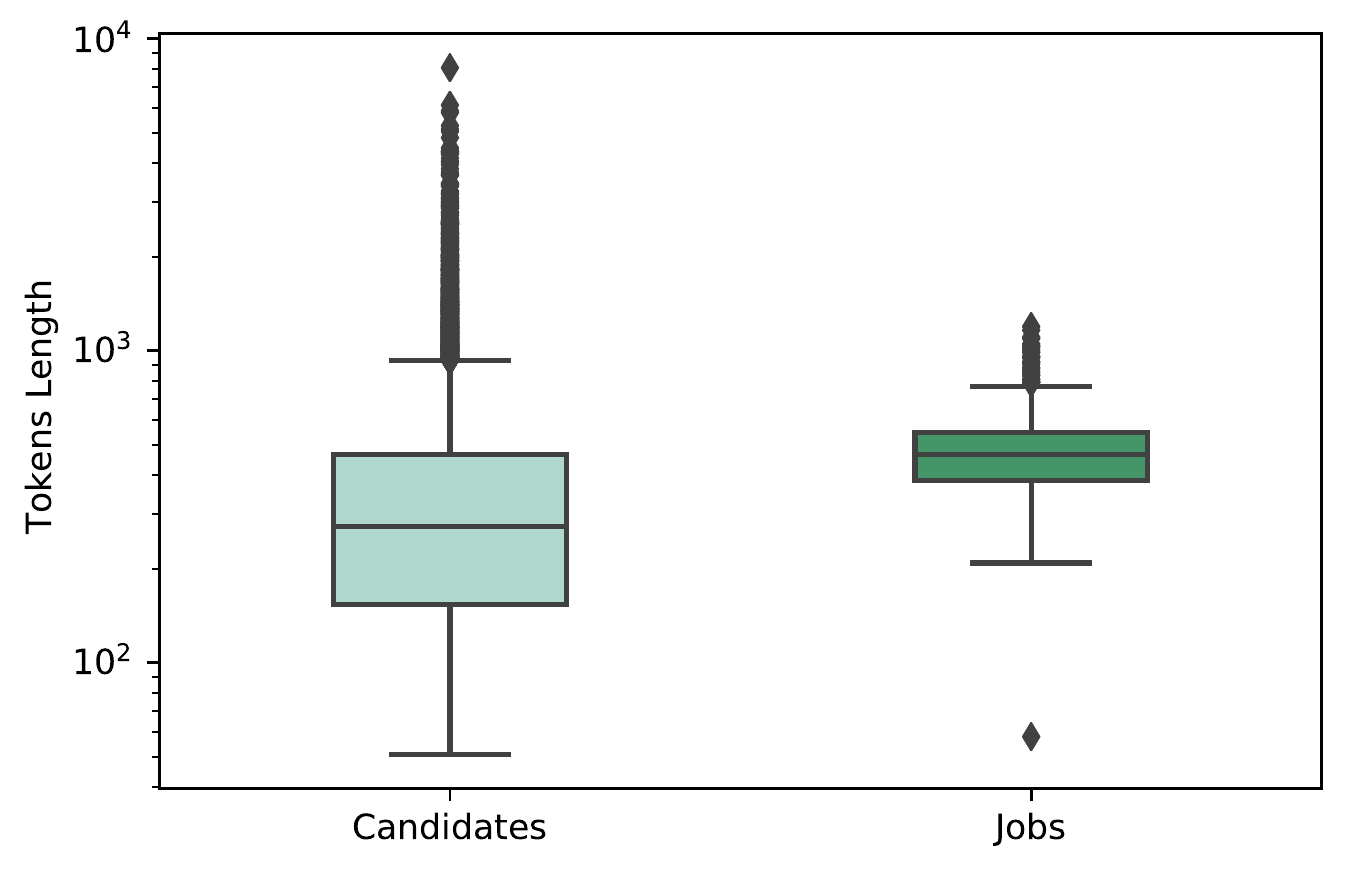}
\caption{The tokens length for the candidates and jobs dataset}
\label{tokens_size}
\end{figure}

Since we have a French dataset, we need a model that was pre-trained in the same language. The CamemBERT was introduced as a state-of-the-art language model that was pre-trained on the French subcorpus \cite{martin2019camembert}. Given the limit size of words a BERT model can support, we plotted the number of tokens of the textual dataset (jobs and candidates). Figure \ref{tokens_size} shows the tokens distribution of the candidates and jobs.

\begin{figure}[H]
\centering
\includegraphics[width=120mm,scale=0.2]{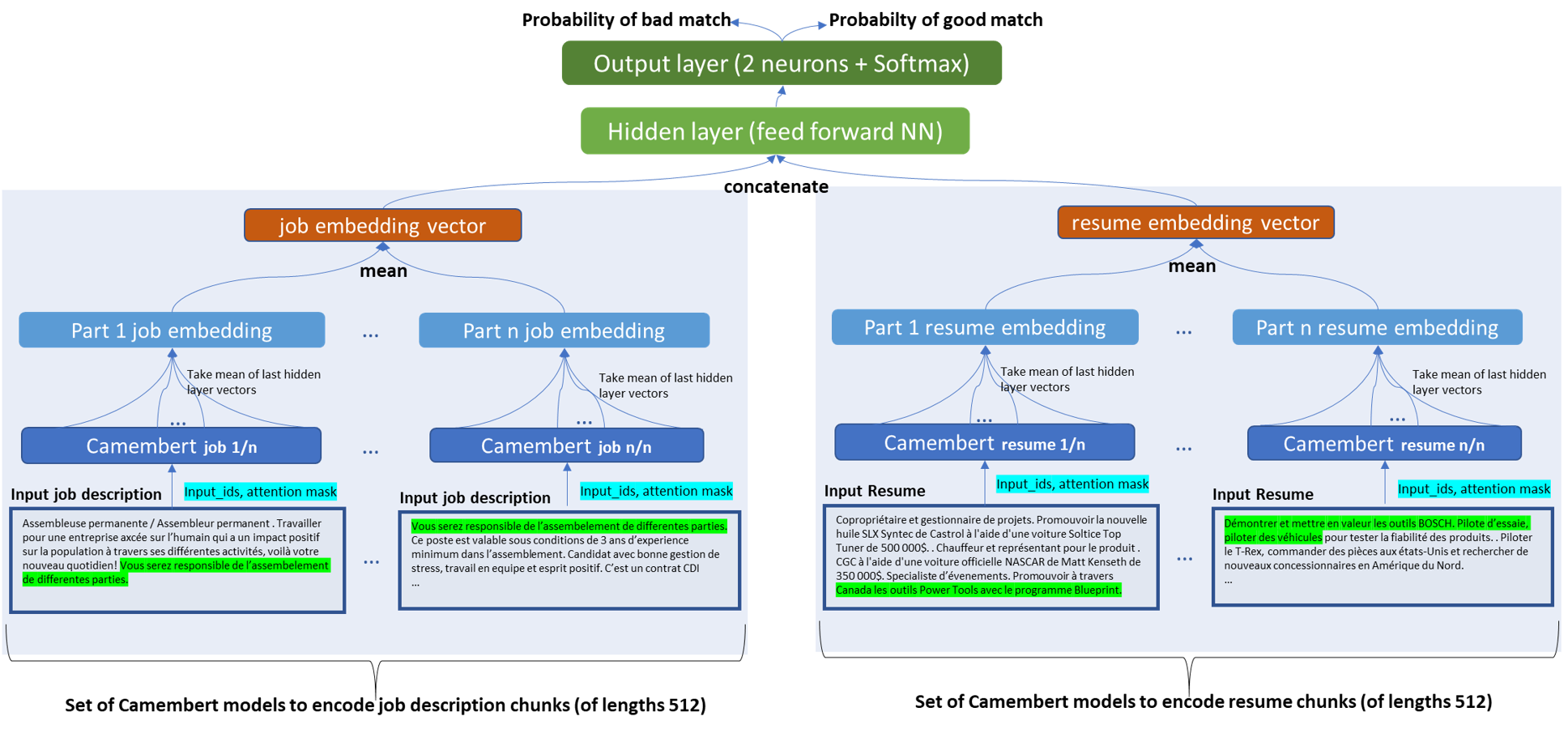}
\caption{Architecture of multiple Camembert architecture}
\label{multi-bert-arch}
\end{figure}

In the proposed architecture, we use multiple Camembert models to encode the input job description and resume. 
Figure~\ref{multi-bert-arch} illustrates the detailed architecture of the model. \vspace{0.1cm}

\noindent
\emph{\textbf{Approach:}}  
We propose a job-resume matching architecture where the number of Camembert models to consider is calculated depending on the training/test dataset (tokens size). We choose the architecture which at most results in losing 10\% (This threshold can be modified) of the input text content. The architecture structure is therefore dynamic and solely depends on the threshold we set and the dataset. By decreasing the loss threshold, we may use a high number of Camembert models and that would result in a high number of parameters which will be difficult to train and converge to the right spot given the data size and computation power. Using a small number of Camembert models will result in loss of information which in turn would affect the performance. It's therefore a trade-off, and the threshold should be chosen wisely.

We keep an overlap of 50 tokens between consecutive chunks in order to prevent context information loss (see text highlighted with green in Figure~\ref{multi-bert-arch}). The final job description/resume representation is given by averaging the representations of each chunk as shown in Equations \ref{eq1}
and \ref{eq2}:
\begin{equation}
job\_embeds = \frac{1}{nbr\_job\_chunks} \sum_{i=1}^{nbr\_job\_chunks} job\_embeds_{i}
\label{eq1}
\end{equation}

\begin{equation}
resume\_embeds = \frac{1}{nbr\_resume\_chunks} \sum_{i=1}^{nbr\_resume\_chunks} resume\_embeds_{i}
\label{eq2}
\end{equation}

The final job description and resume contextualized vectors (i.e., embeddings of length 768) are concatenated (resulting in a vector of length 2*768) and fed to a feed forward neural network for final classification. It’s composed of one hidden layer of 768 neurons (can be tuned later) and ReLU activation function shown in Equation \ref{eq3} to get an intermediary representation of the input of size 768 (i.e., hidden\_output): 
\begin{equation}
hidden\_output = ReLU(W_{768}^t.concatenated\_embeds+b_{768})
\label{eq3}
\end{equation}

The resulting vector representation of the full input is then fed to a final output layer of 2 neurons. The output logits shown in Equation \ref{eq4} represent scores for the ‘match’ and ‘not match’ output classes. 
\begin{equation}
logits = W_{2}^t.hidden\_output+b_{2}
\label{eq4}
\end{equation}

We then apply Softmax function shown in Equation \ref{eq5} to the output logits to get the probabilities of the input job description and resume to ‘match’ or ‘not match’. The output class associated with the highest probability is the predicted class.

\begin{align}
\label{eq5}
\begin{split}
p_{match} = \frac{e^{logits_{1}}}{\sum_{i=0}^{i=1}{e^{logits_{i}}}} 
\\
p_{not\_match} = \frac{e^{logits_{0}}}{\sum_{i=0}^{i=1}{e^{logits_{i}}}}
\end{split}
\end{align}

During training, we finally compute Binary Cross Entropy loss shown in Equation \ref{eq6} and run backpropagation and optimization steps towards minimization of the loss function. During inference we only output the probabilities as the model is already trained. 
\begin{equation}
Bi\_Cross\_Entropy = -q_{match}\log(p_{match}) -q_{not\_match}\log(p_{not\_match})
\label{eq6}
\end{equation}
Where $p$ is the predicted probability and $q$ is the ground truth probability.

\vspace{0.5cm}
\noindent
\emph{\textbf{Results:}}  
We used a stratified split of our initial data into 80\% for training and 20\% for testing. The training set is then split to 80\% for actual training and 20\% for validation.

We set a maximum threshold of data loss of 10\%. The model dynamically chose 1 Camembert for encoding job descriptions (resulting in loss of 4.72\%  of job description data) and 2 Camembert models for encoding resumes (resulting in loss of 6.74\%  of data). With the high number of model parameters, we encounter Out Of Memory issues with batch size higher than 8, on a Tesla T4 GPU (from Google Colab). Therefore, we run training with batches of size 4. Each epoch took around 30min for training and 7min for validation.

Starting from the third Epoch, the model performance did not improve. We therefore stopped at the third epoch and saved the dedicated model. The model achieves 78\% accuracy and an F1-score of 65\%. The full metrics values are reported in Table \ref{tab}. We can conclude that the model is predicting better the non-match label. Thus, additional improvement can be applied \ie using weighted loss function or augmented data, where the best combinations can be validated empirically.

\begin{table}[H]
\centering
\caption{Performance of multiple Camemberts on test set}
\label{tab}
\begin{tabular}{|l|l|c|c|c|}
\hline
 & Label & precision & recall & F1-score \\ \hline
\multirow{2}{*}{\begin{tabular}[c]{@{}l@{}}CamemBERT \\ Windowing\end{tabular}} & not match (0) & 0.8 & 0.93 & 0.86 \\ \cline{2-5} 
 & match (1) & 0.63 & 0.34 & 0.44 \\ \hline
\multirow{2}{*}{\begin{tabular}[c]{@{}l@{}}CamemBert \\ Windowing + Overlap\end{tabular}} & not match (0) & 0.78 & 0.99 & 0.87 \\ \cline{2-5} 
 & match (1) & 0.87 & 0.20 & 0.33 \\ \hline
\end{tabular}
\end{table}

\subsubsection{Features similarity and candidates filtering out}
At this stage of the project, we want to filter out the candidates that don't fit the job exclusive requirements, \ie required experience years.
Using Regex functions, we will determine if there are required years of experience of using a tool, or having specific skills (language, technical hard skills, etc).

\subsubsection{Matching candidates to job offer}
We will create a ranking method based on the matching models. Typically, the JD/Resume matching can be (1) related to text mining and natural language processing techniques such as text classification; (2) sentence matching based on vector similarities ; (3) sentence pair modeling matching; (4) neural network via encoding the JD and resume into a shared space and compute their matching using cosine similarity.


\subsection{Traceability \& Explainability of the matching system}
In this section, we present the possible traceability and explainability ways of the proposed models to the different stakeholders concerned with the matching system.

\subsubsection{Language model Interpretability and Explainability}
After optimization of the model to get the best performing version given our data, we are willing to focus more on the interpretability and explainability aspects of the model.

For that, we aim to run experiments to understand how the model takes decisions to accept/refuse candidates given their resumes and the job description. We will also investigate the parts of resume/job description to which the model pays more attention. This will help understand what the model has learned and its limitations. Which in turn help to tweak the model for better performance.

Recently, the Language Interpretability Tool (LIT)~\cite{lit} was published with an open-source implementation for assessing the explainability of AI models (especially in NLP) through different graphs and functionalities. Using it would be of great help for our use case.

\subsubsection{\RQFour}





At this stage of the project, we will propose explainability reports specifically for the JD/Resume matching we suggested in RQ2. We believe that the better explanation we can have of the matching decision, the more valuable the feature we should have.

The Explainability is about trust. It’s important for the three stakeholders (candidates, job poster, recruiter) of the matching decision to understand its behavior. Depending if the good or bad decisions the model makes, it’s important to have visibility into how they were made. However, the decision models based on deep neural networks behave as black-boxes and fail to provide total explanations for their predictions. The state of the art still advances to provide closer explanations of the model decision, such as the LIME (Local Interpretable Model-Agnostic Explanations) model-agnostic \cite{ribeiro2016should} that has a principle of perturbing the input around its neighborhood and seeing how the model's predictions behave. Similarly, Rationalizing Neural Predictions is trying to predict what part of a paragraph a model is used to make a decision \cite{lei2016rationalizing}.

In our case, we would first like to use such a model-agnostic as a baseline to determine if there are certain sections, keywords, attributes that help the language model make its decision. However, to be able to explain this, we would like to see how the ontology can contribute to better explain a decision of a neuronal network based model. We suppose that a unified ontology for both resume and job description will help us to assign an explanation of the behavior of our language model algorithms. Such experimentation can be validated with qualitative analysis collaborated with the airudi clients (sources of our dataset).

\begin{figure}[H]
\centering
\includegraphics[width=120mm,scale=0.2]{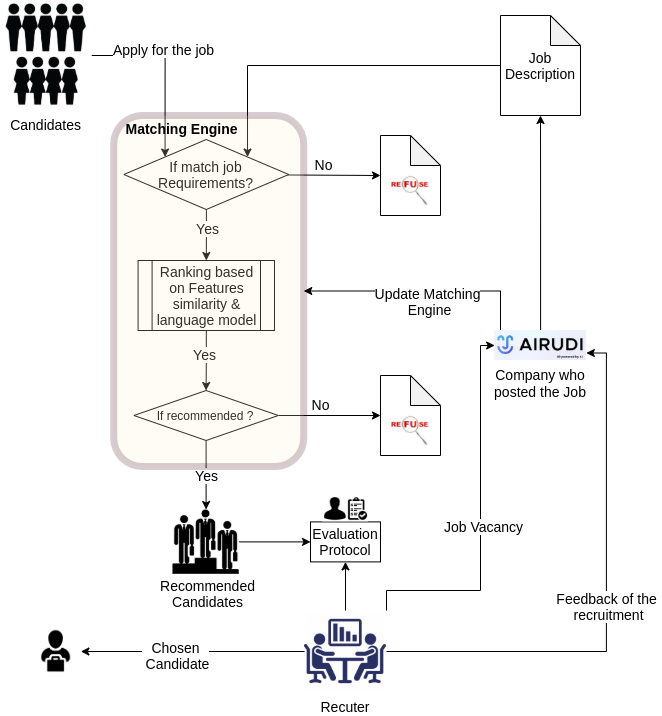}
\caption{Preliminary overview of the proposed explainable system for the concerned stakeholders}
\label{explain}
\end{figure}

We present an overview of the proposed explainable reports we plan to provide to the concerned three stakeholders (candidates, job poster, recruiters) of the matching decisions in Figure \ref{explain}. 

\paragraph{Explanation for a candidates}
The process is enabled, when the \textbf{third-party company} (Airudi) receives a job vacancy from a \textbf{recruiter}. Airudi publishes the job online, and the persons apply for the job, including existing persons already sending their resumes looking for a job vacancy. The matching engine should compute the most eligible candidatures to the posted job, and then prepare an explainable report for the candidates: (1) that are eliminated in the early stage because they do not match the explicit features \ie age, years of experiences, etc. (2) that matched the explicit features, but they have been ranked among the last. A special report will be generated including the details of the trained models and how their skills were not enough to be retained. A comparison between the list of recommended persons for the job with each refused candidate will be made, to include a list of requirements in the report. (3) The persons who were recommended to the recruiter (to be interviewed), but not offered the job will be compared to the chosen candidate, and a similar report indicating the differences characterizing the hired person.       
    
    
     
\paragraph{Explanation for a recruiter}
The recruiter will receive: (1) a list of the recommended candidates from the matching engine, including a highlighted reasons of each person with the reasons that make him matching the job; (2) A features comparison between the candidates will help him to choose easily the most appropriate person needed for the job. Then, the recruiter will evaluate the candidates for the highlighted features that made them eligible for the job.

\paragraph{Explanation for a job poster}
The job poster needs to be updated with the reasons a person was accepted from the proposed list of candidates. The accepted person can be anyone from the recommended list. In that case, a comparison with the candidates will highlight the hidden reason that someone other than the first ranked person was chosen. The company that posted the job will update its recommendation model engine.

\subsubsection{\RQFive}

\begin{figure}[H]
\centering
\includegraphics[width=120mm,scale=0.2]{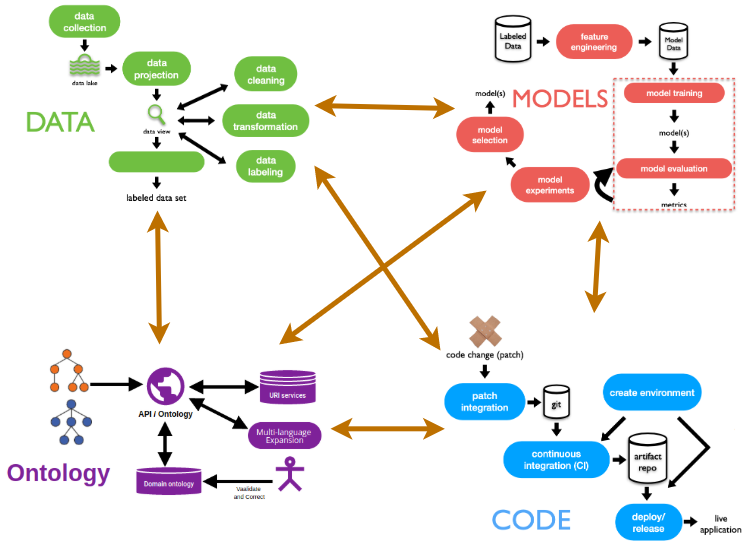}
\caption{The artifacts that should be continuously traceable in the matching JD/Resume environment}
\label{tracability}
\end{figure}





To respond to this research question, there is a need to understand the existing traceability tools in order to choose the most appropriate one(s) that fit our proposed architecture. Subsequently, a comparison and a set of best practice usage of ML traceability tools need to be addressed. The traceability module has to be adapted to include the following requirements as shown in Figure \ref{tracability}:
\begin{itemize}
    \item Tracking continuously the saved dataset with their different evolution from their unstructured sources to include (cleaning, transformation, labeling).
    
    \item Ensure an up-to-date recommendation pipeline evolution, \eg in case a candidate will ask an audit of decisions made by the system. The system should be able to pinpoint exact model version, data set versions used for training, etc.
    \item Updating the different knowledge bases, due to their continuous evolution, once a new feature (skill, education, etc ) appears and needs to be included.
    
    \item Taking in consideration that we have an ontology that can be evolved and can be used particularly with certain models or having transformations (\ie embeddings). Our learning models will be dependent on some parts of the text and other parts of the ontology. We will try to model all system elements that evolve continuously with different compatibility (new ontology release, updated dataset, models' hyperparameters) to be able to trace back and reproduce the system decisions.
    \item Traceability can also be useful for debugging a language model.
\end{itemize}

Tracking data and machine learning in a software repository will introduce new files describing the workflow evolution at the different stages of a project lifetime. This new data may lead to a complexity growth in the case of wrong usage, which opens the door to analyze the co-evolution of such traceability tools with the source code.

We can evaluate the system effectiveness traceability by trying to recover back the system at a random point of its pipeline evolution.

\section{Preliminary results}
We share our preliminary results regarding a case study on open source GitHub projects that are using DVC and how the complexity of the traceability pipelines co-evolution with source code files. Best practice lessons of such tool usage are provided. However, traceability and explainability are two complementary methods in such an environment.

The following study will help us to learn the best practices of integrating such traceability tools in our system. We have this work accepted in \textbf{Saner2021} \cite{amine_saner21}.



\section{Conclusion and Future Work}

An explainable, traceable automated e-recruitment is a need in the cross-border labor market to find the appropriate candidate that matches a proposed job. This project aims to propose an effective e-recruiting tool to suggest the best candidates for the job postings. We will review and provide the State-of-the-art in the JD/Resume matching systems. This study proposes an e-recruiting architecture that considers JD/Resume matching by combining knowledge bases with a pretrained transformer-based machine learning model such as BERT.
Furthermore, the system will generate an explainable report that can be useful for the stakeholders to know the JD/Resume matching system decision. Finally, this research will help to automatize the e-recruitment systems by making them suitable for fair, explainable, and traceable JD/Resume matching.

The on-going SLR presented in the section \ref{method_slr} is the first step to achieve our thesis: \ThesisHyp. We aim to continue this research to achieve the following short and long-term goals which will allow concluding our thesis.

\begin{figure}[H]
\centering
\includegraphics[width=160mm,scale=0.2]{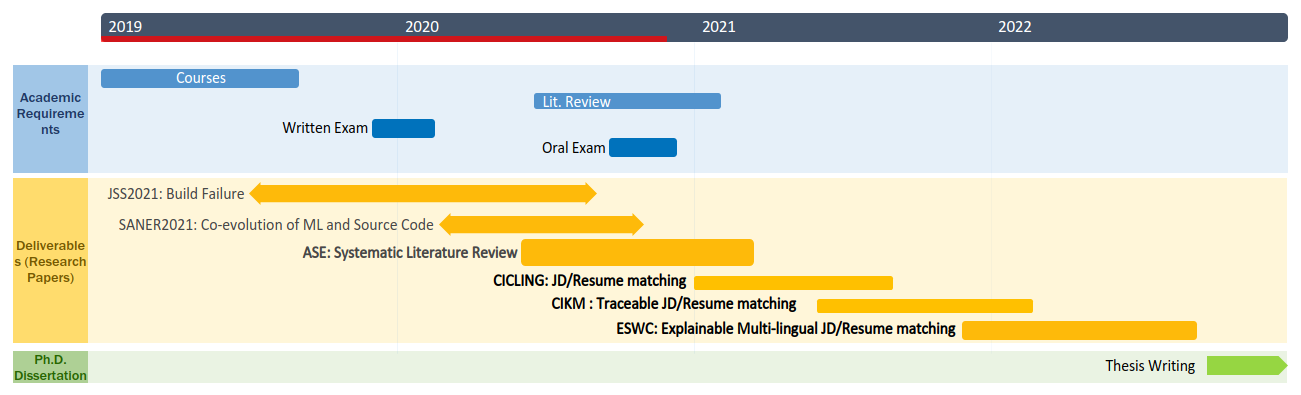}
\caption{Research timeline}
\label{timeline}
\end{figure}

We summarize our thesis research timeline in Figure~\ref{timeline}. It shows the evolution of the research in time during the past and next two years. The academic requirements include course activities, literature review, and the synthesis exam, written and oral part of the research proposal. The deliverable phase includes the preparation for one journal paper and three conference papers. Finally, the Ph.D. dissertation phase includes the Ph.D. thesis preparation and revision and defense. We plan to finish this thesis work by the Winter session of 2022.

All results will be published in Q1 ranked journals and NLP conferences (ASE, CICLING, CIKM, ESWC)


\newpage
\bibliographystyle{ieeetr}
\bibliography{proposal}
\newpage
\appendix

\end{document}